\documentclass[a4paper,11pt,preprint,pre]{revtex4-1}
\usepackage[utf8x]{inputenc}
\usepackage{graphicx}
\usepackage{amsmath}
\usepackage[usenames]{color}
\usepackage{ulem}

\def\refpar[#1]{(\ref{#1})}
\def\msd{{\rm m.s.d.}~}
\newcommand{\beq}{\begin{equation}}
\newcommand{\eeq}{\end{equation}}
\newcommand{\beqa}{\begin{eqnarray}}
\newcommand{\eeqa}{\end{eqnarray}}
\newcommand{\ba}{\begin{array}}
\newcommand{\ea}{\end{array}}
\newcommand{\dd}{\mathrm{d}}

\begin{document}

\title{Single File Diffusion of particles with long ranged interactions: damping and finite size effects}
\author{Jean-Baptiste Delfau} \author{Christophe Coste}\author{Michel Saint Jean}
\affiliation{ Laboratoire MSC. UMR CNRS 7057 et Universit\'e Paris Diderot -Paris7  \\B\^atiment Condorcet 
  10 rue Alice Domon et L\'eonie Duquet  \\  75205  PARIS Cedex 13 -  France
}

\date{\today}

\begin{abstract}
We study the Single File Diffusion (SFD) of a cyclic chain of particles that cannot cross each other, in a thermal bath, with long ranged interactions, and arbitrary damping.  We present simulations that exhibit new behaviors specifically associated to systems of small number of particles and to small damping. In order to understand those results, we present an original analysis based on the decomposition of the particles motion in the normal modes of the chain. Our model explains all dynamic regimes observed in our simulations, and provides convincing estimates of the crossover times between those regimes.

PACS~: 05.40.-a	Fluctuation phenomena, random processes, noise, and Brownian motion
66.10.cg	Mass diffusion, including self-diffusion, mutual diffusion, tracer diffusion, etc. 
47.57.eb	Diffusion and aggregation

\end{abstract}

\maketitle
\thispagestyle{empty}

\section{Introduction}

When Brownian particles are confined along a line in a quasi one dimensional channel so narrow that they cannot cross each other, anomalous diffusion appears and strongly subdiffusive behaviour can be observed. This phenomenon called Single File Diffusion was first noticed in 1955 by Hodgkin and Keynes~\cite{Hodgkin55} who were studying water transport through molecular-sized channels in biological membranes. Since then, SFD also appeared in the diffusion of molecules in porous materials like zeolites~\cite{Gupta95,Hahn96,Chou99},  of charges along polymer chains~\cite{Wang09}, of ions in electrostatic traps~\cite{Seidelin06}, of vortices in band superconductors~\cite{Besseling99,Kokubo04}  and of colloids in nanosized structures~\cite{Wei00,Cui02,Lin02,Lin05,Koppl06,Henseler08} or optically generated channels~\cite{LutzJPC04,LutzPRL04}. Even though SFD can be encountered in a lot of various physical systems, most of the theoretical studies devoted to it are generally restricted to the simplest case~: an infinite overdamped system with hard core interactions.

In this paper, we present simulations results concerning finite systems of long range interacting particles. In particular, we focus on the dependency of the diffusion properties with the number of particles $N$ and the damping coefficient $\gamma$. We exhibit new behaviors, specifically associated to systems of small number of particles and to small damping. In order to interpret those results, we present an original analysis based on the decomposition of the particles motion in the normal modes of the chain.

In the thermodynamic limit (infinite systems with finite density $\rho$), for overdamped dynamics with hard core interactions, several analytical models~\cite{Harris65,Levitt73,vanBeijeren83,Arratia83,DeGennes71} predict that at long times, the mean square displacement of a particle of mass $m$ grows as $F_{H} \sqrt{t}$ with the mobility $F_{H}$ given by 
\beq
F_{H}=\frac{2}{\rho} \sqrt{\frac{D_0}{\pi}} = \frac{2}{\rho} \sqrt{\frac{k_B T}{\pi m \gamma}},
\label{eq:mobhard}
\eeq
with $D_0 = k_B T/( m \gamma)$ the single particle free diffusion coefficient at temperature $T$, and $k_B$ Boltzmann's constant. If the interactions are long-ranged, only two analytical approaches have been undertaken so far~\cite{Kollmann03,Sjogren07}, for overdamped systems in the thermodynamic limit. There it is proven that the m.s.d. grows as $F_S \sqrt{t}$ at long times, with a mobility $F_S$ that depends on the interaction potential, through the isothermal compressibility $\kappa_T$ \cite{Coste10} or the spring constant $K \equiv \rho/\kappa_T$,
\begin{align}
\label{eq:mobform}
F_S = {2\over \rho} S(0,0)\sqrt{{ D_{eff} \over \pi}} = 2 k_B T  \sqrt{\frac{ \kappa_T}{\pi m \gamma  \rho}} = 2 k_B T  \frac{ 1}{\sqrt{\pi m \gamma  K}},
\end{align}
where $S(0,0) \equiv S(q \to 0,t = 0)$ is the long wavelength static structure factor of the particles. The diffusivity $D_{eff}$ is the  effective diffusivity of a Brownian particle, taking into account its interactions with the other Brownian particles \cite{Nagele96}, and differs from the single particle diffusivity $D_0$. In its last version, the expression~\eqref{eq:mobform} can be interpreted by considering that we can derive $F_S$ from $F_H$ if we replace the interparticle distance $1/\rho$ by the mean square displacement $k_B T/K$ of a particle in the potential well due to its neighbors. In appendix~\ref{sec:asympt}, we recover the formula \eqref{eq:mobform} without the assumption of overdamped Langevin dynamics \refpar[langevinbase].

In numerical simulations and experiments, the systems are obviously finite. Periodic boundary conditions are used in simulations, and annular geometries in experiments. As a consequence of finite size effects,  the asymptotic behavior at long time is always $\langle\Delta x^2\rangle  = D_N t$. All particles in the system are then totally correlated and diffuse as a single effective particle of mass $N \times m$ \cite{vanBeijeren83}. We will recover it from our analytical model and provide measurements of the diffusion coefficient $D_N$ in good agreement with this interpretation. The SFD regime may nevertheless be observed in finite systems if the damping and the particles number are high enough, in a manner that will be precised by an analytical approach of finite systems dynamics.

However, most theoretical and experimental studies have been performed for overdamped systems only. This is the initial assumption in the existing models for long range interacting particles \cite{Kollmann03,Sjogren07}. The relevant experiments were generally done with solutions of collo\"ids~\cite{Wei00,Cui02,Lin02,LutzJPC04,LutzPRL04,Lin05,Koppl06,Henseler08} for which overdamping is a safe assumption. The simulations \cite{Herrera07,Herrera08} are shown in \cite{Coste10} to be in good agreement with the theoretical prediction of Kollmann, but they also assume overdamping in the choice of the simulation algorithm. In order to explore \emph{underdamped} systems, we have previously studied the diffusion in a circular channel of millimetric steel balls electrically charged \cite{Coste10}. In this experiment, identical metallic beads are held in a plane horizontal condenser made of a silicon wafer and a glass plate covered with an optically transparent metallic layer. A constant voltage is applied to the electrodes, inducing a charge distribution of the beads. The condenser is fixed on loudspeakers excited with a white noise voltage, and we have checked that this mechanical shaking behaves as an effective thermal bath \cite{Coupier06,Coupier07,Coste10}. In this system the measurement of the damping constant $\gamma$ proved that the balls diffusion is underdamped \cite{Coupier06}. We have observed that the \msd of the particles exhibit the SFD scaling predicted for overdamped systems, with a prefactor that is only slightly higher than the theoretical prediction~\eqref{eq:mobform}. Unfortunately, we were not able to tune the damping constant experimentally. Thus, in order to investigate the specific role of damping on the diffusion of finite systems we have developped numerical simulations that allows easy changes of the damping constant.

The paper is organised as follows~: Section~\ref{sec:simulation}  is devoted to the description of the algorithm used in our numerical simulation. In section~\ref{sec:results}, we present our numerical results and exhibit new behaviors specific to systems of small numbers of underdamped particles. We characterize the various regimes for the \msd scaling with time, and define the crossover times between those regimes. In section~\ref{sec:discussion},  we give a physical interpretation of those regimes in the framework of our analytical model. We recover the various scaling laws for the \msd, and provide estimates of the various crossover times, showing their dependency on the damping and particle numbers. We summarize our results in section~\ref{sec:conclusion}. Two appendices are devoted to complementary calculations.

\section{Description of the simulation}
\label{sec:simulation}

\subsection{A line of particles with long-ranged interactions}
\label{subsec:model}

\begin{figure}[htb]
\centering
\includegraphics[width=7cm]{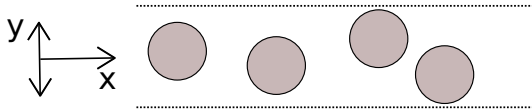}
\caption{\label{schem}Scheme of the system}
\end{figure}

 We consider point particles of mass $m$ in a horizontal plane $(xOy)$, submitted to a thermal bath at temperature $T$. The particles are confined by a quadratic potential in $y$ in such a way that they cannot cross each other, as if they were diffusing in a narrow channel (see Fig.~\ref{schem}). This lateral confinement is chosen to mimic as well as possible experimental situations. We have checked that its strength do not influence the system behavior provided the beads stay ordered.
 
We describe  the dynamics with the Langevin equation.  Let $\mathbf{r}_i = (x_i,y_i)$ be the position of the particle $i$. We do not take into account the gravity, thus describing horizontal systems. The particle is submitted to a confinement force $-\beta y_i \mathbf{e}_y$ of stiffness $\beta$ and to the interaction potential $U(\mathbf{r}_i)$, so that Langevin equation reads
\begin{align}
\label{langevinbase}
 \ddot{\mathbf{r}}_i + \gamma \dot{\mathbf{r}}_i +\frac{\boldsymbol{\nabla} U(\mathbf{r}_i)}{m} + \frac{\beta}{m}y_i\mathbf{e}_y = \frac{\boldsymbol{\mu}(t)}{m}
\end{align}
with $\gamma$ the damping constant and $\boldsymbol{\mu}$ a random force. In our simulations, the random force has the statistical properties of a white gaussian noise. Therefore, its components on both axes must satisfy: 
\begin{align}
\label{eq:wgn1}
\langle\mu_x(t) \rangle = 0,\quad  \langle\mu_y(t) \rangle = 0,\quad \langle\mu_x(t) \mu_y(t')\rangle = 0,\\
\label{eq:wgn2}
\langle \mu_x(t)\mu_x(t') \rangle = \langle \mu_y(t)\mu_y(t') \rangle =  2 k_B T m \gamma \delta (t-t') \end{align}
where $k_B$ is Boltzmann's constant and $\langle\cdot\rangle$ means statistical averaging.

It is suitable to put those equations in dimensionless form, defining the following dimensionless variables: $t=\widetilde{t}/\gamma$ and $x=\widetilde{x}\sqrt{{k_B T}/({m \gamma^2})}$. It gives us
\begin{align}
 \ddot{\widetilde{\mathbf{r}_i}} + \dot{\widetilde{\mathbf{r}_i}} + \widetilde{\boldsymbol{\nabla}} \widetilde{U}(\widetilde{\mathbf{r}_i}) + \frac{\beta}{m\gamma^2}\widetilde{y_i}\mathbf{e}_y =  \widetilde{\boldsymbol{\mu}}(\widetilde{t})
 \label{eq:langevinadim}
\end{align}
with the dimensionless quantities:
\begin{align}
\widetilde{U}(\widetilde{\mathbf{r}_i})=\frac{U(\widetilde{\mathbf{r}_i})}{k_B T} , \qquad
\widetilde{\boldsymbol{\mu}}(\widetilde{t})=\frac{\boldsymbol{\mu}(\widetilde{t})}{\sqrt{k_B T m \gamma^2}} 
\end{align}
and the only nonzero correlation (\ref{eq:wgn2}) now reads
\begin{align}
\label{eq:wgn4nodim}
\langle \widetilde{\mu}_x(\widetilde{t})\widetilde{\mu}_x(\widetilde{t'}) \rangle = \langle \widetilde{\mu}_y(\widetilde{t})\widetilde{\mu}_y(\widetilde{t'}) \rangle =   2 \delta (\widetilde{t}-\widetilde{t'}).
\end{align}
For the sake of simplicity, we drop the "tildes" $\widetilde{\quad}$ in the rest of this section.

In order to allow a direct comparison between simulations and experiments, we take the same interaction potential as in our experimental set-up  \cite{Coupier06,Coupier07,Coste10}. It reads
\begin{align}
 U(\mathbf{r}_i)= {U_0} \sum_{j\ne i}K_0\left({\left\vert\mathbf{r}_i-\mathbf{r}_j\right\vert \over \lambda}\right), 
 \label{eq:gammabeads}
\end{align}
where $K_0$ is the modified Bessel function of second order and index $0$, $\lambda$ and $U_0$ two constants. In principle, the sum extends to all particles, but in practice the summation is limited to the first five neighbours of each particle, which ensures a relative precision better than $10^{-7}$ and reduces the calculation time.

\begin{figure}
\centering
\includegraphics[width=6cm]{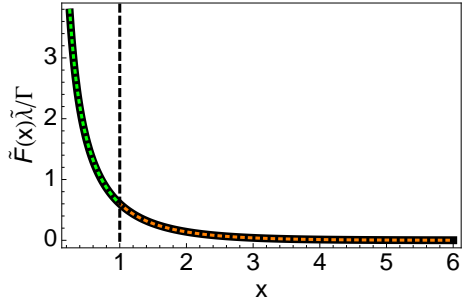}
\caption{\label{pot}Force approximation. The thick black line represents the actual force derived from eqn.~\eqref{eq:gammabeads}, the green and orange dotted lines are respectively the logarithmic and exponential approximations in eqn.~\eqref{eq:forceapprox}.}
\end{figure}

To decrease the computation time further, we replace the Bessel functions in the expression of the force ${\displaystyle \mathbf{F}({\mathbf{r}_i})=-{\boldsymbol{\nabla}} {U}({\mathbf{r}_i}) = \sum_{j \neq i} F_{ij}(\left\vert\mathbf{r}_i-\mathbf{r}_j\right\vert){\mathbf{r}_i-\mathbf{r}_j\over \left\vert\mathbf{r}_i-\mathbf{r}_j\right\vert}}$ by asymptotic expressions,
\beq
\left\{
\begin{array}{l l l}
{\displaystyle{F}_{ij}(x)} & ={\displaystyle\frac{U_0}{{\lambda }} \left[-\frac{1}{x}+bx+cx\ln(x)\right]} & \mbox{for }x<1\\
\\
{\displaystyle{F}_{ij}(x)} & = {\displaystyle\frac{U_0}{{\lambda }} \left[\sqrt{\frac{\pi}{2 x}} e^{-x}\left(1+\frac{a}{x}\right) \right]} & \mbox{for }x>1
\end{array}
\right.
\label{eq:forceapprox}
\eeq
where $a$, $b$, and $c$ are constants. They are chosen in such a way that the force and its derivative are continuous for $x=1$, and that the force is equal to its actual value at this point. Those two approximations fit very well the actual force (see Fig.~\ref{pot}).

\subsection{Algorithm}
\label{sec:algorithm}

The simulation is based on the Gillespie algorithm~\cite{Gillespie96PRE,Gillespie96AJP} that allows a consistent time discretization of the Langevin equation (\ref{eq:langevinadim}). We introduce a time step value ${\Delta t}$, which for consistency has to be much smaller than any other characteristic time-scale of the system. In dimensionless units $\Delta t = 10^{-3}$. Then the velocities $\dot{{x_i}}({t}+{\Delta t})$ and $\dot{{y_i}}({t}+{\Delta t})$ are calculated from updating formula derived from (\ref{eq:langevinadim}), 
\beq
\left\{
\begin{array}{l}
 \dot{{x_i}}({t} +{\Delta t}) = \dot{{x_i}}({t}) - \bigl[\dot{{x_i}}({t}) + \boldsymbol{\nabla} {U}(r_i(t))\cdot\mathbf{e}_x\bigr]\times{\Delta t} + \sqrt{2 {\Delta t}} \times {\mu_x}({t})\\
\\
 {\displaystyle\dot{{y_i}}({t} + {\Delta t}) = \dot{{y_i}}({t}) - \left[\dot{{y_i}}({t}) + \frac{\beta}{m \gamma^2} {y_i}({t})+ \boldsymbol{\nabla} {U}(r_i(t))\cdot\mathbf{e}_y\right]\times{\Delta t} + \sqrt{2 {\Delta t}} \times {\mu_y}({t})}
\end{array}
\right.
\label{eq:langevinnum}
\eeq
where $r_i(t) = \sqrt{x_i(t)^2 + y_i(t)^2}$. The positions ${x_i}({t}+ {\Delta t})$ and ${y_i}({t}+ {\Delta t})$ of all the particles are then calculated from
\beq
{x_i}({t}+{\Delta t})={x_i}({t})+ \dot{{x}_i}({t}){\Delta t}, \qquad {y_i}({t}+{\Delta t})={y_i}({t})+ \dot{{y}_i}({t}){\Delta t}.
\label{eq:langevinposnum}
\eeq
The components of the random noise ${\mu_y}$ and ${\mu_x}$ are sampled in such a way that they have the properties given by equations~(\ref{eq:wgn1}) and~(\ref{eq:wgn4nodim}), hence that they are unit normal random numbers. 

We simulate systems of $N$ particles, with periodic boundary conditions. We get from \refpar[eq:langevinposnum] $N$ equivalent trajectories, because all beads play the same role. The system is simulated during a dimensionless time of $10^3$, which means $10^6$ time-steps. The quantity of interest is the mean square displacement (m.s.d.) along the $x$ direction,
\beq
\langle\Delta x^2(t)\rangle = \langle \left[x(t + t_0) - x(t_0) - \langle x(t + t_0) - x(t_0)\rangle\right]^2\rangle.
\label{eq:defmsd}
\eeq
where $t_0$ is an arbitrary initial time. The ensemble averaging is done on every beads, since they all play an equivalent role. Moreover, the phenomenon is assumed to be stationary, so that $\Delta x^2(t)$ do not depend on $t_0$. For a given time $t$, it makes thus sense to average on the initial time $t_0$. Let $n$ be the overall number of time-steps in one simulation, and $n_t = t/\Delta t$. Then the averaging on the initial time $t_0$ reads
\beq
\langle\langle\Delta x^2(t)\rangle_e\rangle_0 = \sum\limits_{i = 0}^{n - n_t}{\left\{x[(n_t+i)\Delta t] - x(i \Delta t)\right\}^2 \over n - n_t + 1} -\left( \sum\limits_{i = 0}^{n - n_t}{x[(n_t+i)\Delta t] - x(i \Delta t) \over n - n_t + 1}\right)^2,
\label{eq:calcavg}
\eeq
where the index $i$ is such that $t_0 = i \Delta t$, $\langle \cdot \rangle_e$ means ensemble averaging and $\langle \cdot \rangle_0$ means averaging on the initial time $t_0$. This way of averaging greatly improves the statistics when $n_t$ is smaller than $n$. We will use it henceforward, denoting it with the simplified notation $\langle \cdot \rangle$ except in appendix~\ref{sec:average} where it is specifically discussed.

\subsection{Orders of magnitude of the various parameters.}

In our simulations, we work at densities $\rho = 33$, $100$ and $533$ particles per meter. The temperature and interaction strength are such that $\Gamma$ ranges as in experiments \cite{Wei00,LutzPRL04,LutzJPC04,Coste10} and numerical simulations \cite{Herrera07,Nelissen07,Herrera08}. The interest of the simulations is to get acces to parameter values that are difficult or impossible to obtain experimentally. We vary the particles number $N$ between 32 and 1024. This last value is comparable to some simulations \cite{Herrera07,Herrera08} but much greater than in experiments \cite{Wei00,LutzPRL04,LutzJPC04,Coste10}. We vary the damping constant $\gamma$ between 0.1~s$^{-1}$ and 60~s$^{-1}$, extending the experimental range toward small values of $\gamma$. This is to be compared to the cut-off frequency of the chain (see section~\ref{sec:theory}). With our damping constant range, we get access to both the overdamped and underdamped dynamics of the particles, and are thus able to exhibit the subtle behaviors linked to underdamping.

\bigskip

We simulate the same system that was experimentally studied in \cite{Coste10}. In the experiments, the beads number $N$ vary between 12 and 37, and the density $\rho$ is 477, 566 or 654 particles per meter. The mean interparticle distance is thus such that $1.53<1/\rho < 2.10$~mm, to be compared to the range $\lambda = 0.48$~mm of the potential. The dimensionless potential energy is such that $6 < \Gamma < 55$. The damping constant $\gamma$ ranges between 10~s$^{-1}$ and 30~s$^{-1}$ (see \cite{Coupier06}, Fig.~6). For the experimental values of density and potential energy,  the cut-off frequency ranges between $21$~s$^{-1}$ and 37~s$^{-1}$. Experimentally, we are thus in the underdamped regime, as was already noticed in \cite{Coste10}.

\section{SFD of finite systems~: the different regimes}
\label{sec:results}

\begin{figure}[!ht]
\centering
\includegraphics[width=7cm]{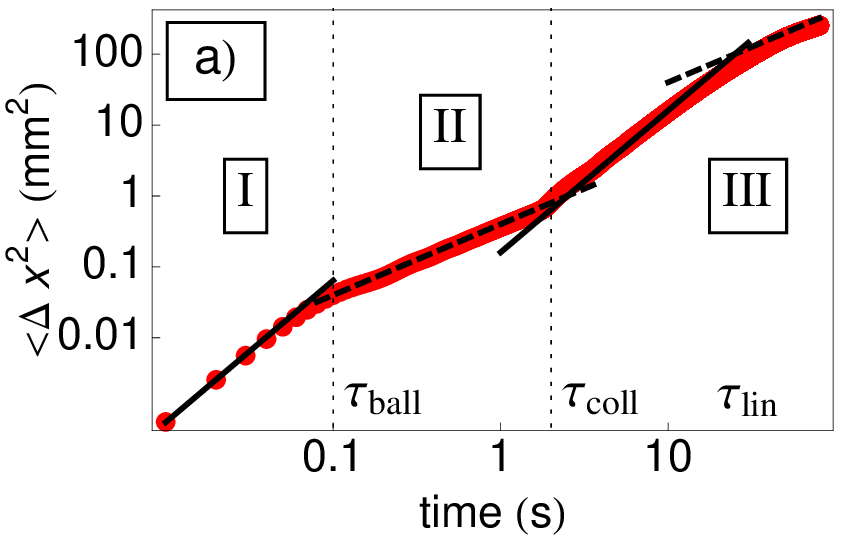}
\includegraphics[width=7cm]{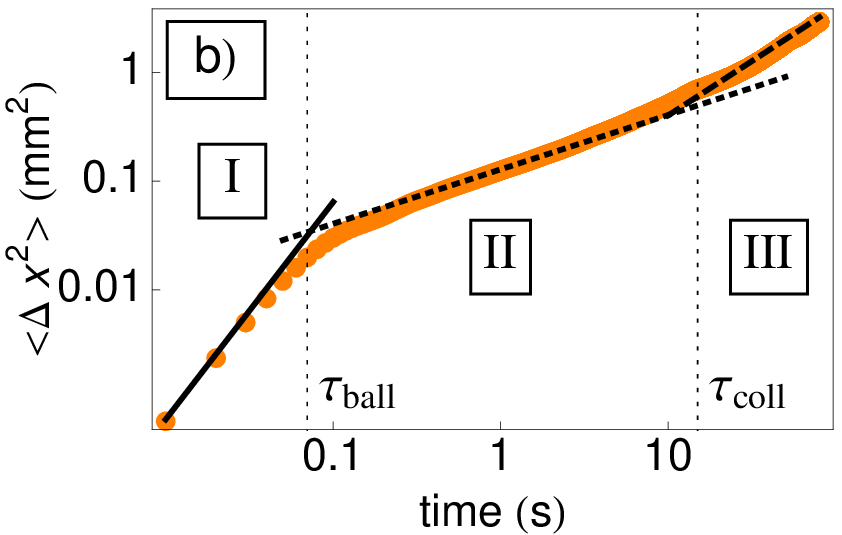}
\caption{\label{regimes}Evolution of the m.s.d. (in mm$^2$) according to the time (in s.) for a chain of 32 particles with density 533~m$^{-1}$, temperature $T=10^{12}$ K and interaction potential $\Gamma \approx 7$. The solid line scales as $t^2$, the dashed line scales as $t$, the dotted line scales as $t^{1/2}$.\\
(a) Damping constant $\gamma = 0.1$~s$^{-1}$. In this low damping case, regime II is characterized by a $t$ scaling, regime III by a $t^2$ then a $t$ scaling.\\
(b) Damping constant $\gamma = 60$~s$^{-1}$. In this strong damping case, regime II is characterized by a $\sqrt{t}$ scaling, regime III by a $t$ scaling.}
\end{figure}

In this section, we present our results about the evolution of the mean square displacement (m.s.d.) $\langle\Delta x^2(t)\rangle$ as a function of the time $t$, and focus on the effects of the particles number $N$ (at fixed density) and on the damping constant $\gamma$. Two typical examples are provided by Fig.~\ref{regimes}. The evolution of the m.s.d. may be described by power laws $\langle\Delta x^2(t)\rangle \propto t^\alpha$, with an exponent $\alpha$ that depends on the observation time. The interpretation that will be detailed in part IV allow us to regroup them into three different regimes~:
\begin{itemize}
 \item during the first regime I, $0 \leq t \leq \tau_{\rm ball}$, the m.s.d. grows according to $H_1 t^2$;
 
\item during regime II,  $\tau_{\rm ball} \leq t \leq \tau_{\rm coll}$, $\langle\Delta x^2\rangle$ may be proportional to $D t$ only, $F_S \sqrt{t}$ only or to  $D t$ then $F_S \sqrt{t}$,  depending on the parameters of the simulation. The coefficient $D$ is \emph{not necessarily} the free diffusion constant $D_0$. When both scalings $D t$ and $F_S \sqrt{t}$ are observed, we define the crossover time $\tau_{\rm sub}$ between them.

 \item A final regime III takes place for $\tau_{\rm coll} \leq t$, $\langle\Delta x^2\rangle = D_N t$ at long times with $D_N \neq D$ and $D_N \neq D_0$. This final asymptotic behavior is sometimes preceded by the scaling $H_Nt^2$ with $H_N \neq H_1$, the crossover time being noted $\tau_{\rm lin}$ .
 
\end{itemize}

\begin{figure}[!ht]
\centering
\includegraphics[width=6cm]{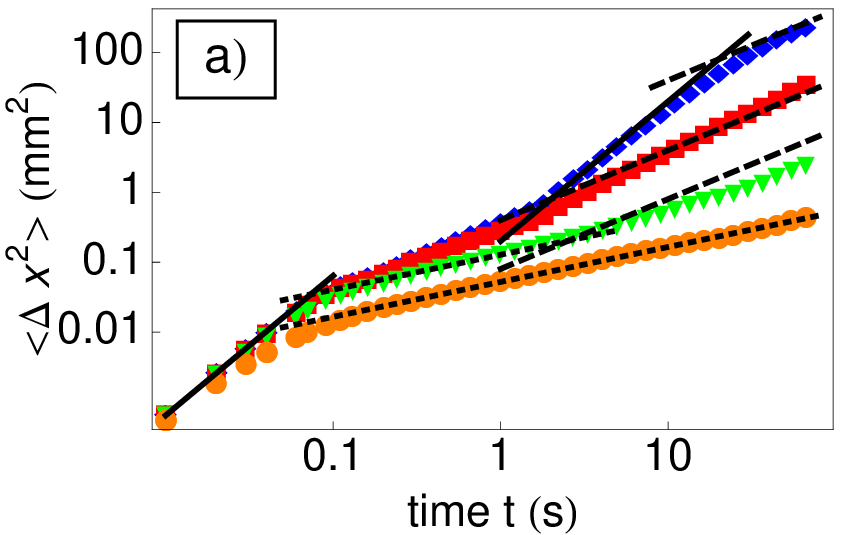}
\includegraphics[width=6cm]{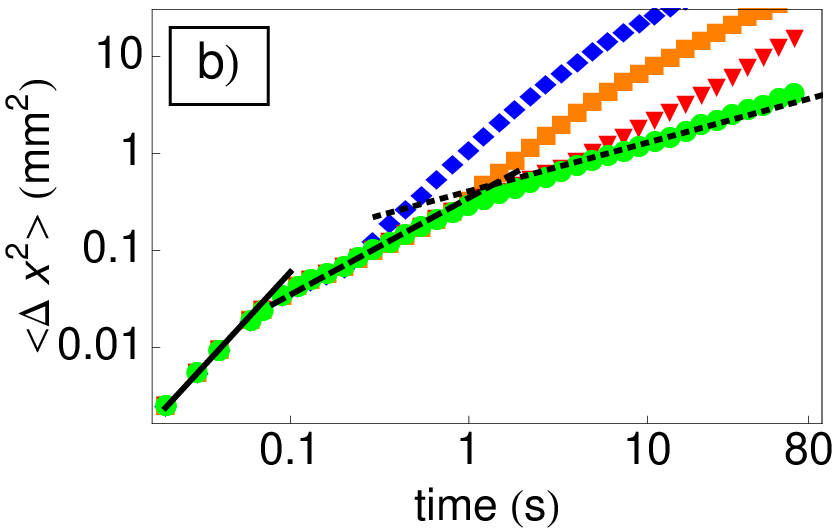}
\includegraphics[width=6cm]{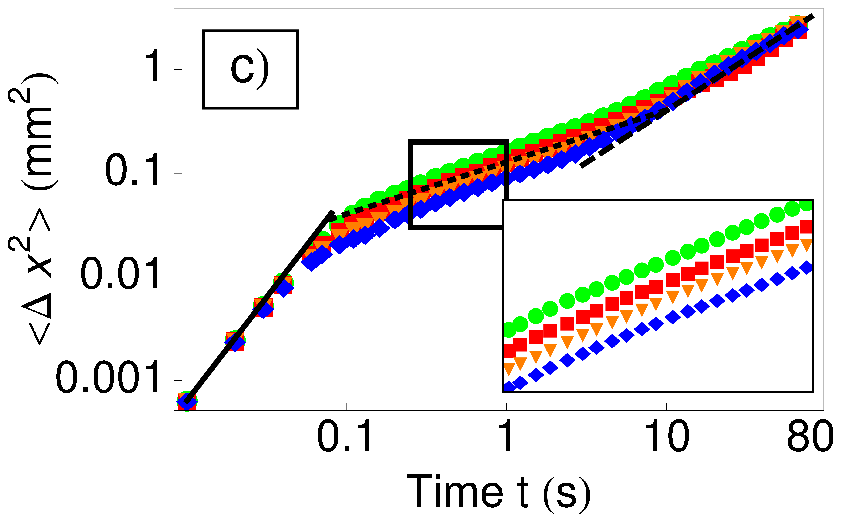}
\includegraphics[width=6cm]{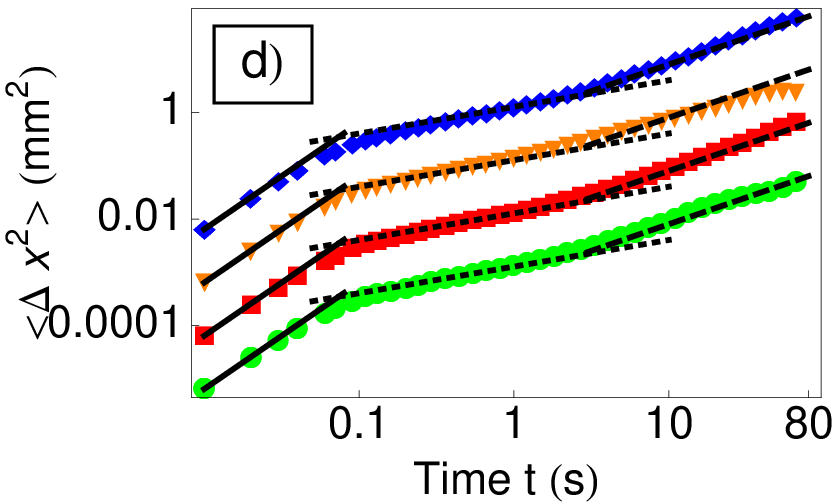}
\caption{\label{variances} Plot of $\langle\Delta x(t)^2 \rangle$ (in mm$^2$) according to the time (in s.) for a density  $\rho = 533$~m$^{-1}$. Unless otherwise specified, the parameters are $N = 32$, $T=10^{12}$~K, $\Gamma \approx 6.8$ and $\gamma = 10$~s$^{-1}$. Specific values are as follows~: (a) $\gamma=0.1\mbox{, }1\mbox{, }10\mbox{ and }60\mbox{ s}^{-1}$ (blue, red, green and orange respectively). (b) $\gamma=1 \mbox{ s}^{-1}$ and $N=4\mbox{, }16\mbox{, }64\mbox{ and }128$ (blue, orange, red and green respectively). (c) $\Gamma \approx 4.4\mbox{, }6.8\mbox{, }9.8\mbox{ and }13.4$ (green, red, orange and blue respectively). (d) $T=10^{10}\mbox{, }10^{11}\mbox{, }10^{12}\mbox{ and }10^{13}$~K, (green, red, orange and blue respectively). The black thick line is~\eqref{eq:solcorreltpscourt}, the dashed line is $F_S \sqrt{t}$ with the mobility $F_S$ given by \eqref{eq:mobform}
 and the dotted line is either $D_N t$ with $D_N$ given by \eqref{cdiffN}, in (a), (c) and (d) or  $D t$ with $D$ given by \eqref{eq:diffeffect} in (b). There are no free parameters in the calculations.}
\end{figure}

\subsection{The small time regime (regime I)}
\label{sec:regimeIdata}

Regime I is defined by an evolution $\langle\Delta x^2\rangle = H_1 t^2$. It is observed in all data displayed in Fig.\ref{variances}, and the prefactor $H_1$ is independent on the damping constant [see Fig.\ref{variances}--a)] ,  on the system size [see Fig.\ref{variances}--b)] and on the interaction potential $\Gamma$ [see Fig.\ref{variances}--c)]. In this time range ($0 \leq t \leq \tau_{\rm ball}$), each particle behaves independently from the others and ensures a ballistic flight at its thermal velocity $\sqrt{k_B T/m}$, so that the constant $H_1$ should thus be equal to $k_B T/m$. The duration of this first \emph{ballistic regime} is called $\tau_{\rm ball}$. From  our data summarized in Fig.\ref{variances}, we measure the constant $H_1$ and show in Fig.~\ref{prefactors}--a) that it is indeed in perfect agreement with its predicted value. 

This behavior is obviously not observed in the simulations of the overdamped Langevin equation \cite{Herrera07,Herrera08,Henseler08,Centres10}, but as already been seen in simulations of the full dynamics \cite{Taloni08}. In \cite{Nelissen07}, they simulate the full Langevin equation but they do not display data for a sufficiently small time to observe the $t^2$ scaling. 

\subsection{The intermediate time regime (regime II)}
\label{sec:regimeIIdata}

If we consider now the second regime, two different behaviors with distinct power laws can be observed~: Fig.~\ref{variances}--a) shows that for the highest values of $\gamma$, $\langle\Delta x^2\rangle$ only grows as $\sqrt{t}$. When $\gamma$ is decreased, a linear evolution in $D t$ appears for $\tau_{\rm ball} < t <\tau_{\rm sub}$. For the lowest values $\gamma = 1$~s$^{-1}$ and $\gamma = 0.1$~s$^{-1}$, the  $\sqrt{t}$ scaling completely disappears. This is clearly a \emph{finite size effect}, as seen in Fig.~\ref{variances}--b). The data displayed in this picture are recorded at a low value $\gamma = 1$~s$^{-1}$, and the $\sqrt{t}$ scaling is not recovered until large numbers of particles, typically $N> 128$. Data from simulations with 256, 512 and 1024 particles (at constant density) superimpose exactly on the data for 128 particles.

We could be tempted to explain the linear evolution in $D t$ by arguing that we observe the diffusion of a free particle, that needs a finite time to feel the effect of confinement. If this should be the case, the diffusion coefficient $D$ should be the diffusion constant for a free particle, which is:
\begin{align}
\label{cdiff}
 D_0=\frac{k_B T}{m \gamma}
\end{align}
In Fig.~\ref{prefactors}--b), we compare our numerical values of $D$ to $D_0$. It is obvious that $D$ is very different from $D_0$ except at the lowest values of the density $\rho$ (that is, low interactions). We shall see in section~\ref{sec:regimeIIdiscussion} that for high interactions the coefficient $D$ actually results from a collective behavior of the particles. In our model [see eqn.~\eqref{eq:appvariancelowgamma}], when $\rho$ is high (high interactions), the coefficient $D$ doesn't depend upon $\gamma$ and is given by
\begin{align}
\label{eq:diffeffect}
 D = \frac{k_B T}{2 \pi} \sqrt{\frac{\kappa_T}{m \rho}} = \frac{k_B T }{2 \pi\sqrt{m  K}}.
\end{align}
In Fig.~\ref{prefactors}--c), we see that at high density the coefficient $D$ is actually a function of $k_B T/\sqrt{m K}$, but with a numerical coefficient that is rather equal to $1/2$. The dependency of $D$ on either the spring constant $K = U''(1/\rho)$ or the compressibility $\kappa_T = \rho/K$ indicates that collective phenomenons are responsible of this behavior, and that the "free particle" hypothesis does not account for the linear behavior observed in strongly interacting systems. The modified Bessel function $K_0$ which gives the behavior of the potential $U(1/\rho)$ [see~\eqref{eq:gammabeads}] is a very quickly increasing function of the density, which explains why this behavior is typical of high density systems.
 
When the subdiffusive regime $\langle\Delta x^2\rangle=F\sqrt{t}$ is observed, as in Fig.~\ref{variances}--c) and Fig.~\ref{variances}--d), we may measure the mobility $F$. As seen in  Fig.~\ref{prefactors}--d), our numerical data are in excellent agreement with the expression $F_S$ given in formula \refpar[eq:mobform], even if our system is underdamped. We discuss in section~\ref{sec:regimeIIdiscussion} the relevance of this equality for \emph{finite} systems.

The numerical data displayed in Fig.~\ref{variances}--c) and Fig.~\ref{variances}--d) are calculated for parameters values that are very close (in particular, the system sizes are equal) to the relevant parameters of the experiments reported in \cite{Coste10}. One can check (see Fig.~5 of \cite{Coste10}) that the value of the \msd is the same. The duration of the experiments is unsufficient to see the final $D_N t$ scaling described in the next section.

\subsection{The large time regime (regime III)}
\label{sec:regimeIIIdata}

\begin{figure}[!ht]
\centering
\includegraphics[width=6cm]{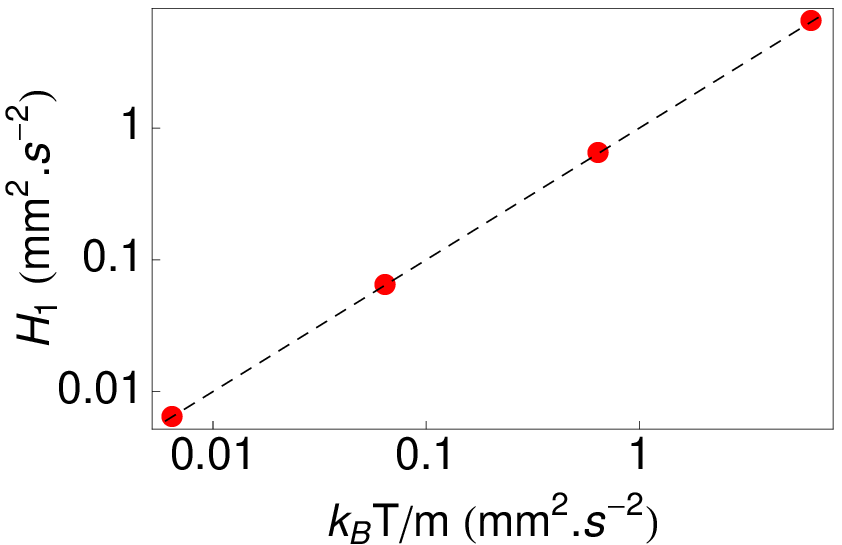}
\includegraphics[width=6cm]{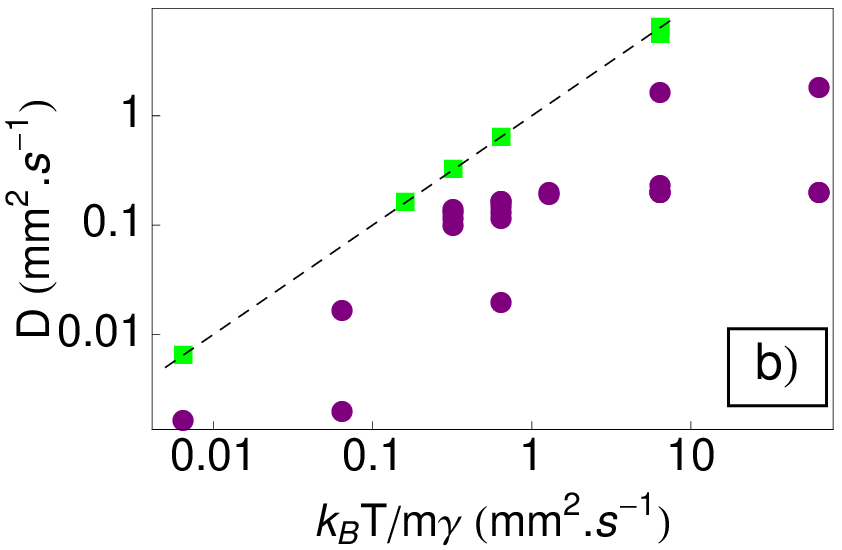}
\includegraphics[width=6cm]{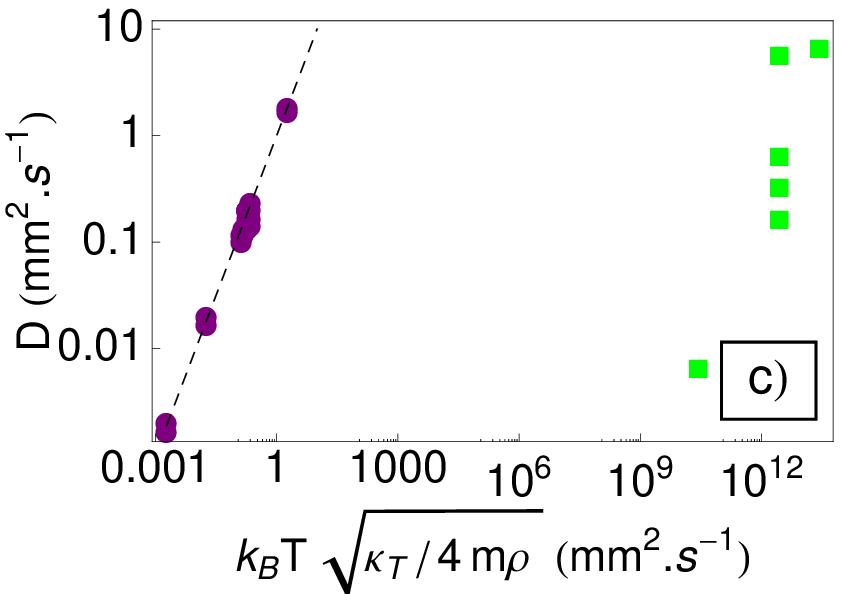}
\includegraphics[width=6cm]{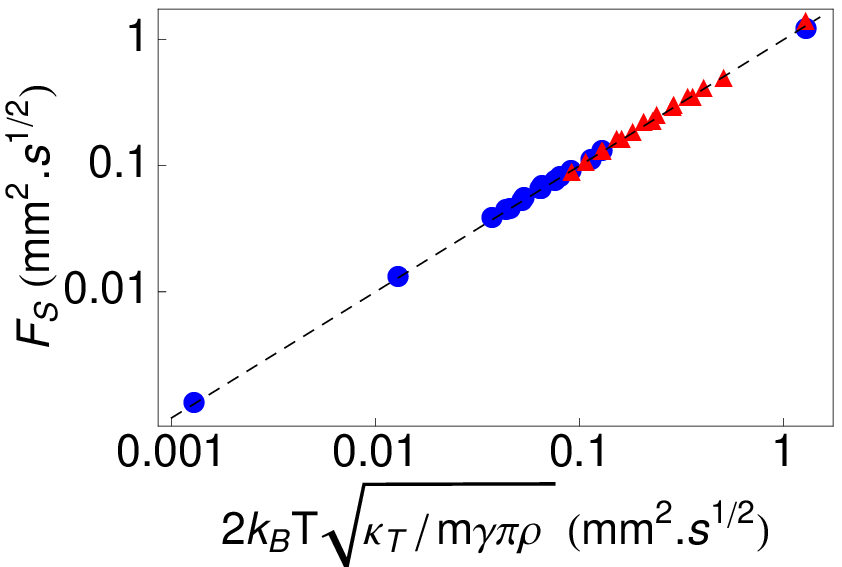}
\includegraphics[width=6cm]{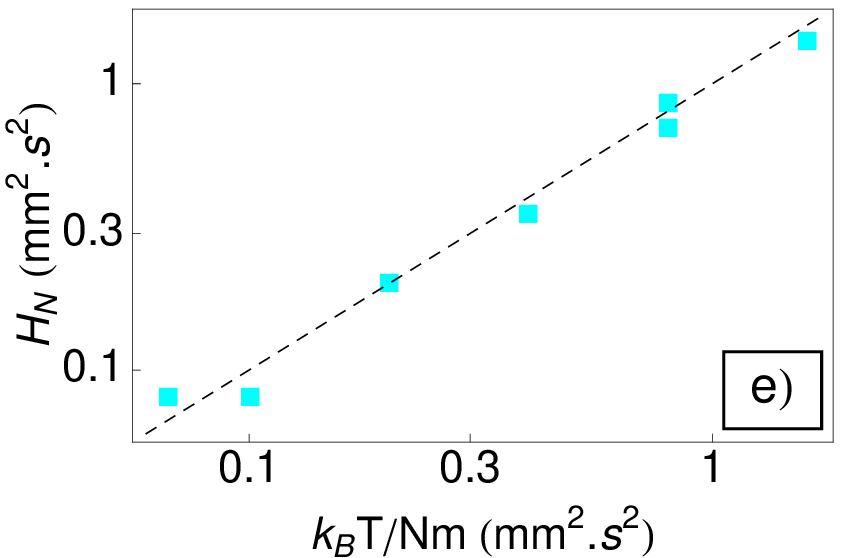}
\includegraphics[width=6cm]{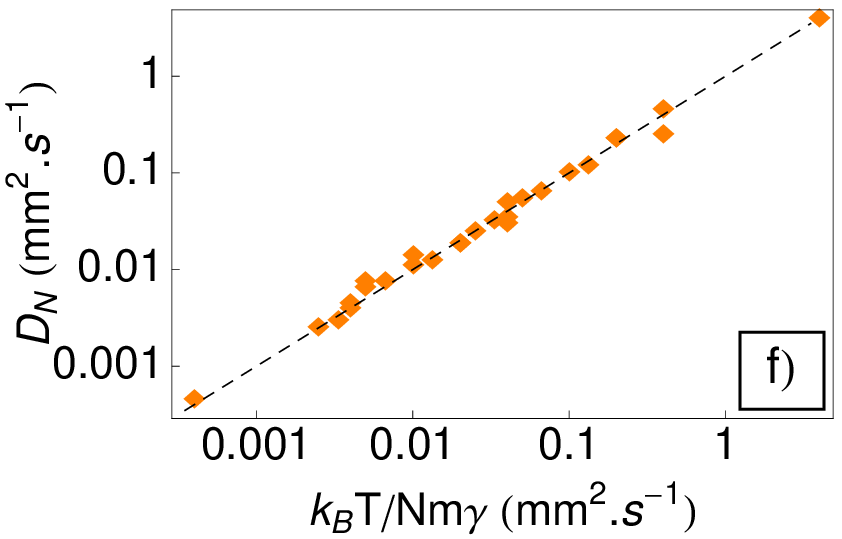}
\caption{\label{prefactors}a) Coefficient $H_1$ as a function of $k_B T/m$.\\
b) Coefficient of diffusion $D$ according to $k_BT/(m \gamma)$ [see~\eqref{cdiff}].  The green squares represent $D$ for systems of low densities ($\rho \approx 100 \mbox{ and } 33 \mbox{ m}^{-1}$) and the purple circles for systems with higher densities ($\rho \approx 533 \mbox{ m}^{-1}$).\\
c) Coefficient of diffusion $D$ according to $k_B T\sqrt{\kappa_T}/(2 \sqrt{m \rho})$ [the numerical coefficient is slightly different from that of~\eqref{eq:diffeffect}]. The green squares represent $D$ for systems of low densities ($\rho \approx 100 \mbox{ and } 33 \mbox{ m}^{-1}$) and the purple circles for systems with higher densities ($\rho \approx 533 \mbox{ m}^{-1}$)\\
d) Mobility $F_S$ according to $2 k_B T\sqrt{\kappa_T}/\sqrt{\pi m \gamma \rho}$ [see~\eqref{eq:mobform}]. The blue correspond to overdamped systems and the red triangles  to underdamped systems.\\
e) Coefficient $H_N$ according to $k_BT/(N m)$ [see~\eqref{cballN}].\\
f) Coefficient of diffusion $D_N$ according to $k_B T/(N \gamma)$ [see~\eqref{cdiffN}].\\
All axes in mm$^2\cdot$s$^{-1}$ and logarithmic scales. All dotted lines are of slope 1.}
\end{figure}

It is quite intuitive that, for very long times and finite systems, all particles become fully correlated and behaves as a single effective particle of mass $N \times m$. It is a property of the translationally invariant mode (see section~\ref{sec:theory}). For small values of $\gamma$ and $N$, the m.s.d. grows according to $H_N t^2$, with a prefactor $H_N$ that is different from the constant $H_1$ introduced in section~\ref{sec:regimeIdata} above. $H_N$ should thus be given by the resolution of the Langevin equation for a free particle of mass $N \times m$~:
\begin{align}
\label{cballN}
 H_N=\frac{k_B T}{N m},
\end{align}
which is in very good agreement with our simulations as shown by Fig.~\ref{prefactors}--e). 

At higher values of $\gamma$, and for larger systems, we only observe a linear evolution of the \msd  $\langle\Delta x^2\rangle = D_N t$ as shown by Fig.~\ref{variances}--c) and \ref{variances}--d). Similarly, $D_N$ should be given by the resolution of the Langevin equation for a free particle of mass $N \times m$:
\begin{align}
\label{cdiffN}
 D_N=\frac{k_B T}{N m \gamma}.
\end{align}
This expression is in very good agreement with our results Fig~\ref{prefactors}--f). 

The interpretation of section~\ref{sec:regimeIIIdiscussion} will show that this regime is dominated by the \emph{collective} behavior of the particles, so that we call $\tau_{\rm coll}$ the time at which these collective behavior takes place. We recover the values of $D_N$ and $H_N$ in section~\ref{sec:regimeIIIdiscussion} from our analytical solution \eqref{eq:solcorrel}, and give an estimate showing that this quadratic scaling is favored by small damping constants and small particles number, as is the case in Fig.~\ref{variances}--a) and \ref{variances}--b).

\subsection{Crossover times}
\label{sec:crossovertimes}

Now that we know the evolution of $\langle\Delta x(t)^2\rangle$ in the different regimes , we can estimate the different crossover times. In this section, we will proceed heuristically, defining the crossover times by requiring continuity of the \msd for successive scalings.

\begin{figure}[!ht]
\centering
\includegraphics[width=6cm]{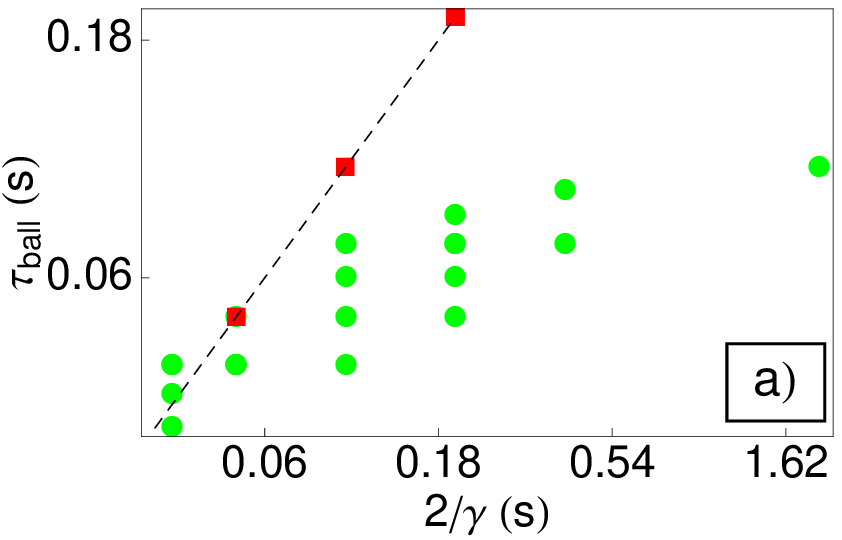}
\includegraphics[width=6cm]{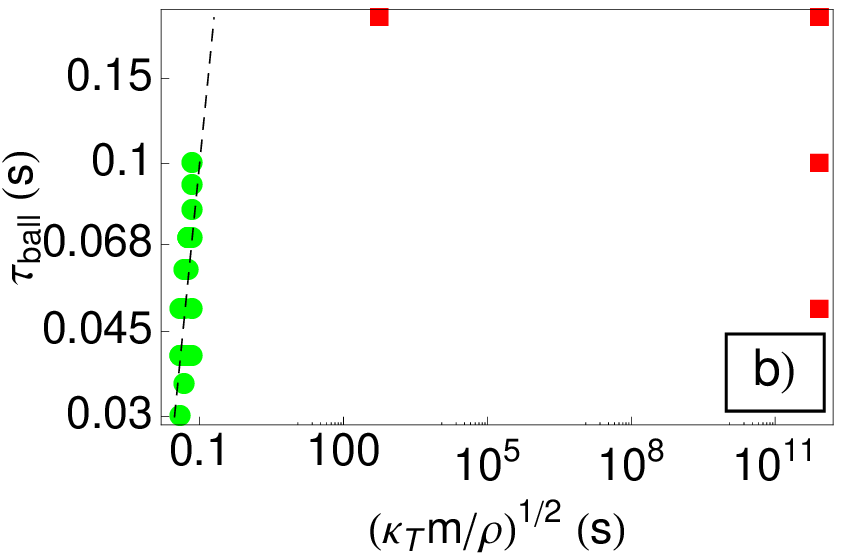}
\includegraphics[width=6cm]{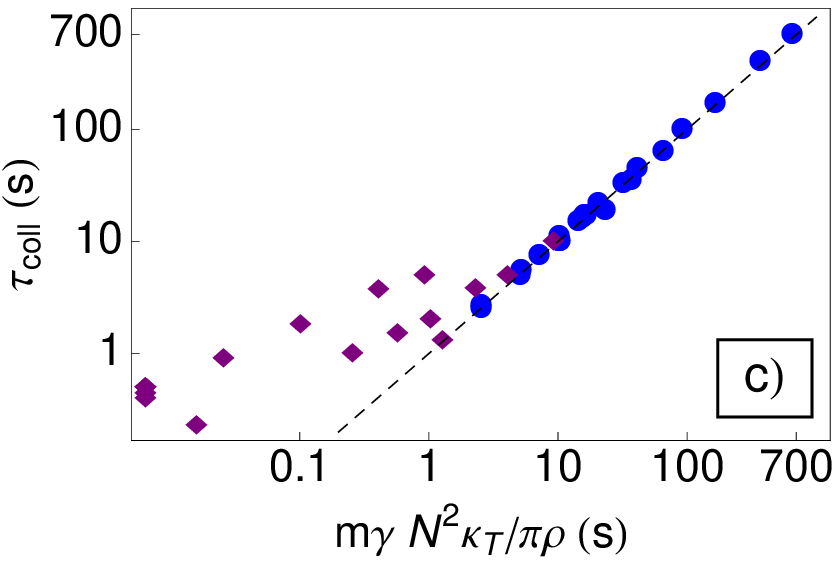}
\includegraphics[width=6cm]{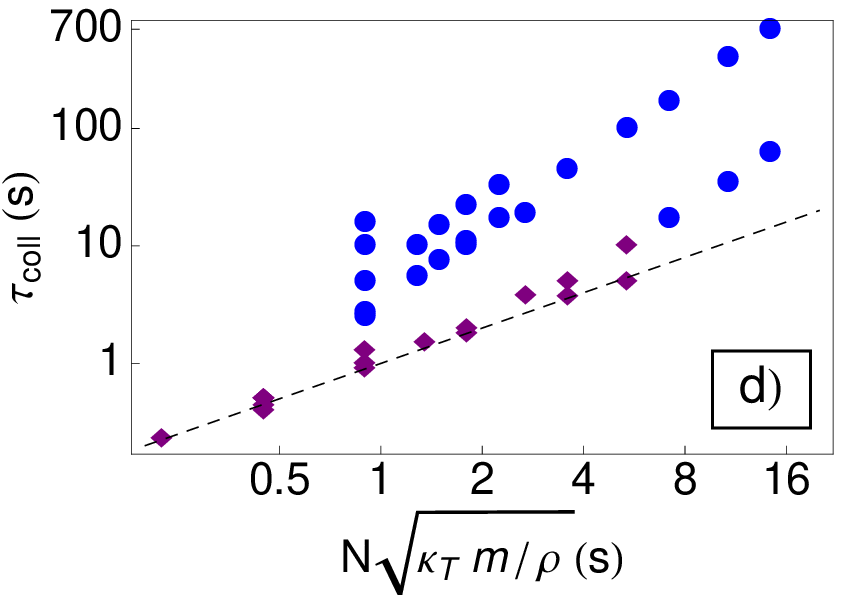}
\caption{\label{tlin}a) Measures of the transition time $\tau_{\rm ball}$ according to $2/\gamma$ [Log scale; see~\eqref{t1weak}]. The green circles represent $\tau_{\rm ball}$ for systems of high densities ($\rho \approx 533 \mbox{ m}^{-1}$) and the red squares for systems with lower densities ($\rho \approx 100 \mbox{ and } 33 \mbox{ m}^{-1}$)\\
b) Measures of the transition time $\tau_{\rm ball}$ according to $\sqrt{\kappa_T m/\rho}$ [Log scale; see~\eqref{t1strong}]. The green circles represent $\tau_{\rm ball}$ for systems of high densities ($\rho \approx 533 \mbox{ m}^{-1}$) and the red squares for systems with lower densities ($\rho \approx 100 \mbox{ and } 33 \mbox{ m}^{-1}$)\\
c) Measures of the transition time $\tau_{\rm coll}$ according to $m \gamma N^2 \kappa_T/\pi \rho$ [Log scale; see~\eqref{t2form}]. The blue circles represent $\tau_{\rm coll}$ for systems with $(N/2\pi)\sqrt{\gamma^2 m \kappa_T/\rho}>1$ and the purple diamonds for systems with $(N/2\pi)\sqrt{\gamma^2 m \kappa_T/\rho}<1$ (see~\eqref{eq:criterium}).\\
d) Measures of the transition time $\tau_{\rm coll}$ according to $N\sqrt{\kappa_T m/\rho}$ [Log scale; the numerical coefficient is slightly different from that of~\eqref{t2form2}]. The blue circles represent $\tau_{\rm coll}$ for systems with $(N/2\pi)\sqrt{\gamma^2 m \kappa_T/\rho}>1$ and the purple diamonds for systems with $(N/2\pi)\sqrt{\gamma^2 m \kappa_T/\rho}<1$ (see~\eqref{eq:criterium}).\\
All axes in s. All dashed lines are of slope 1.}
\end{figure}

Following this method, the ballistic time $\tau_{\rm ball}$ will be the time for which the curves of equation $H_1 t^2$ and $2Dt$ will intersect. Depending on the expression of the diffusion coefficient $D$, we obtain:

\begin{align}
\frac{k_B T}{m} \tau_{\rm ball}^2 \sim 2\frac{k_B T}{m \gamma}\tau_{\rm ball} \quad \Longrightarrow \quad
\tau_{\rm ball} \sim \frac{2}{\gamma}
\label{t1weak}
\end{align}
for weakly interacting systems. For strongly interacting systems, one has to considere  the effective diffusion coefficient $D$ of \eqref{eq:diffeffect}, which gives
\begin{align}
\frac{k_B T}{m} \tau_{\rm ball}^2 \sim \frac{k_B T}{\pi} \sqrt{\frac{\kappa_T}{m\rho}}\tau_{\rm ball} \quad \Longrightarrow \quad
\tau_{\rm ball} \sim \frac{1}{\pi}\sqrt{\frac{\kappa_T m}{\rho}}.
\label{t1strong}
\end{align}
We performed measurements of $\tau_{\rm ball}$ for different values of parameters and reported them on Fig.~\ref{tlin}--a) and Fig.~\ref{tlin}--b). One can clearly see that two different mecanisms are at stake: the green circles which represent systems of high densities ($\rho \approx 533 \mbox{ m}^{-1}$), which are associated to strongly interacting particles, can be easily distinguished from the red squares which represent systems with lower densities ($\rho \approx 100 \mbox{ and } 33 \mbox{ m}^{-1}$), thus weakly interacting particles. Formula~\eqref{t1weak} seems in good agreement with the transition times of weakly interacting systems whereas formula~\eqref{t1strong} fits the values of $\tau_{\rm ball}$ for strongly interacting ones.

Let us now consider $\tau_{\rm coll}$.  For overdamped and large systems, it will be the time for which the curves of equation $F_S\sqrt{t}$ and $D_{N}t$ intersect, thus giving
\begin{align}
F_S \sqrt{\tau_{\rm coll}} &= 2D_N\tau_{\rm coll} \quad \Longrightarrow \quad \tau_{\rm coll} = \frac{F^2}{4D_N^2} = \frac{S(0,0)^2}{D_0^2} \times D_{eff} \times \frac{N^2}{\pi \rho^2}= \frac{m \gamma N^2 \kappa_T}{\pi \rho}.
 \label{t2form}
\end{align}
Note that starting from equation~\eqref{eq:mobform}, using the fact that $D_{eff}=D_0/S(0,0)$, introducing the length $L = N/\rho$ of the chain, we may recast this expression to obtain
\begin{equation}
\tau_{\rm coll} = \frac{L^2}{\pi D_{eff}}
\end{equation}
which is interesting as it tells us that $\tau_{\rm coll}$ can be seen as the time necessary for a given particle to diffuse over the length of the system $L$, with the effective diffusion coefficient that takes into account its interactions with the other particles. 

In the case of small damping and small systems, the SFD behavior is not observed in the intermediate regime [see Fig.~\ref{variances}--a) and Fig.~\ref{variances}--b)], being replaced by a $D t$ scaling with $D$ given by \eqref{eq:diffeffect}, and the collective regime begins by a $t^2$ evolution [see~\ref{sec:regimeIIIdata} and \eqref{cballN}]. The time $\tau_{\rm coll}$ may thus be estimated by
\begin{align}
2 D \tau_{\rm coll} = H_N\tau_{\rm coll}^2\quad \Longrightarrow \quad \frac{k_B T}{\pi \sqrt{m K}}\tau_{\rm coll} = \frac{k_B T}{N m}\tau_{\rm coll}^2 \quad \Longrightarrow \quad \tau_{\rm coll} = \frac{N}{\pi}\sqrt{\frac{m}{K}}.
\label{t2form2}
\end{align}

One can see on figure~\ref{tlin}--c) that $\tau_{\rm coll}$ is in very good agreement with equations~\eqref{t2form} for large values of $\gamma N^2$. For small values of  $\gamma N^2$, the data rather follow~\eqref{t2form2}, according to the analysis provided in section~\ref{sec:regimeIIIdiscussion} and particularly~\eqref{eq:tpslongbis}. 

\section{Theoretical analysis}
\label{sec:discussion}
\subsection{A chain of springs and point masses in a thermal bath}
\label{sec:theory}

In order to analyse the results presented in the last section, we have studied the Langevin dynamics of a chain of $N$ beads of mass $m$, aligned along the $x$ axis, interacting with a pair potential $U(x)$, with  nearest neighbors interactions. Those two simplifying assumptions, a strictly 1D system  with nearest neighbors interactions only, are as we will see in excellent agreement with the actual dynamics. 

Small oscillations around the equilibrium position are  described by linear springs of force constant $K = U''(1/\rho)$ where $\rho$ is the particle density at equilibrium \cite{Brillouin53}. Let $x(l,t)$ be the position of particle $l$ at time $t$. The equation of motion reads
\beq
{\dd^2  \over \dd t^2}x(l,t) = -\gamma{\dd  \over \dd t}x(l,t) + {K \over m}\left[x(l+1,t) - 2x(l,t) + x(l-1,t)\right] + \frac{{\mu}(l,t)}{m},
\label{eq:chain}
\eeq
with the same notations as in \S~\ref{subsec:model}. Let's consider a chain with periodic boundary conditions. We may introduce the discrete Fourier transform
\beq
X(q,t) = \sum\limits_{l = 1}^N e^{i q l}x(l,t), \qquad x(l,t) = \frac{1}{N}\sum\limits_{k = 1}^N e^{-i q_k l}X(q_k,t), 
\label{eq:Fourier}
\eeq
with $q_k = -\pi + {2\pi k / N}$ for $k = 1, \ldots, N$. From now on, we simplify the notations, dropping the dependency of the modes $q_k$ on the natural number $k$ and replacing summations on $k$ by summations on $q$. The variance of the displacement $x$ may be calculated from the Fourier modes $X(q,t)$, as $\langle \Delta x^2 (t)\rangle = \sum_q \langle \Delta X^2(q,t) \rangle/N^2$ with
\beq
 \langle \Delta X^2(q,t) \rangle \equiv \left\langle \bigl[X(q,t) - \langle X(q,t)\rangle\bigr]\bigl[X(-q,t) - \langle X(-q,t)\rangle\bigr]\right\rangle.
\label{eq:defrmsFourier}
\eeq

Let us first considere the mode $q = 0$, that will be noted simply $X(t)$. It follows the equation
\beq
{\dd^2  \over \dd t^2}X(t) +\gamma{\dd  \over \dd t}X(t)  =  \frac{{\mu}(q = 0,t)}{m}.
\label{eq:modezero}
\eeq
Physically, this is the equation for a free particle of mass $m$ in a thermal bath at temperature $T$, with damping constant $\gamma$. The solution is composed of two parts, $X_d(t)$ which corresponds to the deterministic motion of the particle, and the fluctuating part $X_\mu(t)$ which depends linearly on the random forcing $\mu(q = 0,t)$. It reads
\beq
X(t) - X^0 = \frac{\dot X^0}{\gamma}\left[1 - e^{-\gamma t}\right] + \frac{1}{m}\int\limits_0^t \dd t' \int\limits_0^{t'} \dd t'' e^{-\gamma(t' - t'')}\mu(q = 0,t''),
\label{eq:modezerosolu}
\eeq
where $X^0 \equiv X(t = 0)$ and $\dot X^0 \equiv \dot X (t = 0)$ are the initial conditions. The corresponding mean square displacement contribution measured in the simulations is the double average  defined by \refpar[eq:calcavg]. It is calculated in the appendix~\ref{sec:average} and reads
\beq
\langle \Delta X^2 \rangle  =  2\frac{N k_B T}{m \gamma}\left[t  - \frac{1}{\gamma}\left(1 -  e^{-\gamma t}\right)\right].
\label{eq:doubleavgzero}
\eeq
\bigskip

We now considere the modes $q \neq 0$. Using the periodic boundary conditions $x(l,t) = x(l+N,t)$, we see that each mode $X(q,t)$ of wave number $q\neq 0$ follows the equation
\beq
{\dd^2  \over \dd t^2}X(q,t) +\gamma{\dd  \over \dd t}X(q,t) + {2K \over m}(1 - \cos q)X(q,t)  =  \frac{{\mu}(q,t)}{m}.
\label{eq:mode}
\eeq
The roots of the characteristic polynomial associated to this equation are
\beq
\omega_\pm(q) \equiv -{\gamma \over 2} \pm \sqrt{{\gamma^2 \over 4} - \omega_q^2},\qquad \omega_q^2 \equiv 2{K \over m}(1 - \cos q).
\label{eq:defomega}
\eeq
Physically, the modes $X(q,t)$ behaves as a particle of mass $m$ in an harmonic potential well with pulsation $\omega_q$, forced by the random force  $\mu(q,t)$. The solution of \refpar[eq:mode] is readily obtained as
\beq
X(q,t) = \frac{\dot X(q,0) + \omega_-(q)X(q,0) }{ \omega_+(q) - \omega_-(q)}e^{\omega_+(q)t} + \frac{\omega_+(q)X(q,0)  - \dot X(q,0) }{ \omega_+(q) - \omega_-(q)}e^{\omega_-(q)t} + X_\mu(q,t).
\label{eq:solcorrelnonnul}
\eeq

The  translationally invariant mode $q = 0$ scales as $t^2$ at small time $t \leq 1/\gamma$, then scales as $t$ when $t \gg 1/\gamma$. The modes with nonzero wave number scales as $k_B T t^2/(N m)$ at small time, and saturate toward the constant value $2 k_B T/(N m \omega_q^2)$ at very large time. 

\begin{figure}[htb]
\centering
\includegraphics[width=6.5cm]{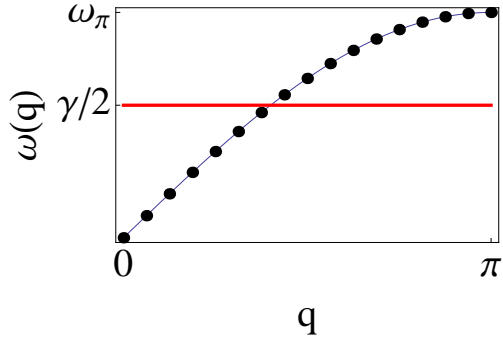}
\caption{\label{fig:reldisp} Dispersion relation $\omega(q)$ as a function of $q$, for $q \geq 0$ (the curve is obviously symmetric for $q \leq 0$). The continuous line is valid for the infinite chain, the dots represent the modes for $N = 32$. The modes such that their frequency is less than $\gamma/2$ are overdamped, the other ones are underdamped.}
\end{figure}

At intermediate time, the behavior of the modes $q \neq 0$ is determined by the relative values of $\omega_q$ and $\gamma/2$, as illustrated in Fig.\ref{fig:reldisp}.\\
-- The modes such that $\omega_q < \gamma/2$ are overdamped. They reach their saturation value at a time $t_{\rm sat} \sim 1/\vert \omega_+(q)\vert$. The shortest saturation time is associated to $q = \pi$, and reads $t_{\rm sat}^{\rm min} \sim \gamma/\omega_{\pi}^2$ if $\omega_\pi <\gamma/2$ or $t_{\rm sat}^{\rm min} \sim 2/\gamma$ otherwise.\\
-- The modes such that $\omega_q > \gamma/2$ are underdamped. They oscillate at a frequency $\omega(q) \equiv \sqrt{\omega_q^2 - \gamma^2/4}$. Below a time which is roughly  $1/\omega(\pi)$, all underdamped modes scale as $t^2$.\\
Finally, all modes $q \neq 0$ saturate toward $2 k_B T/(N m \omega_q^2)$ at large time.

We do not distinguish the overdamped ($\gamma > 2\omega_q$) and the underdamped ($\gamma < 2\omega_q$) modes, because the final result \refpar[eq:flucqnonnul] is in both cases a real function with the same formal expression. The averaging process is explained in appendix~\ref{sec:average}. One gets
\beq
\langle \Delta X^2(q,t) \rangle  = {2 N k_B T \over m  \omega_q^2}\left[1 + {\omega_-(q)e^{\omega_+(q)t} \over \omega_+(q) - \omega_-(q)} - {\omega_+(q)e^{\omega_-(q)t} \over \omega_+(q) - \omega_-(q)}\right].
\label{eq:flucqnonnul}
\eeq
We remark that the limit $q \to 0$ is not singular, and by taking it properly in this expression one recovers \refpar[eq:doubleavgzero]. We think that it is nevertheless physically convenient to distinguish between the translationally invariant mode $q = 0$ and the others, because they lead to different asymptotic behaviors.

Using \refpar[eq:flucqnonnul] together with \refpar[eq:doubleavgzero], we obtain the mean square displacement as
\beq
\langle \Delta x^2 (t)\rangle = {2 k_B T\over N m}\left\{\frac{t}{\gamma}  - \frac{1}{\gamma^2}\left(1 -  e^{-\gamma t}\right) + \sum\limits_{q\neq 0} {1 \over \omega_q^2}\left[1 + {\omega_-e^{\omega_+t} \over \omega_+ - \omega_-} - {\omega_+e^{\omega_-t} \over \omega_+ - \omega_-}\right]\right\},
\label{eq:solcorrel}
\eeq
which will be the basis for the following discussion. 

In the limit of very large damping, that is in the absence of the inertial term, a solution of \refpar[eq:chain] have been provided by L. Sj\"ogren \cite{Sjogren07}. We have thus extended his calculations to underdamped systems~\footnote{More precisely  ${\protect\langle} \Delta x^2 (t){\protect\rangle}$ is equal to the correlation $C(0,t)$ defined by Sj{\protect\"o}gren  \cite{Sjogren07}, and to $2 W(t)$ introduced by Kollmann \cite{Kollmann03}. This is easily seen from the definition~\eqref{eq:defmsd}, because with the double averaging on the initial conditions and on the random noise [see appendix~\ref{sec:average} and eqns.~\eqref{eq:modezerosolu} and \eqref{eq:solcorrelnonnul}] we get ${\protect\langle}{\protect\langle} x(t + t_0) - x(t_0){\protect\rangle}_e{\protect\rangle}_0 = 0$, and then stationarity ensures ${\protect\langle}{\protect\langle}[x(t + t_0) - x(t_0)]^2{\protect\rangle}{\protect\rangle} = {\protect\langle}{\protect\langle} [x(t ) - x(0)]^2{\protect\rangle}{\protect\rangle}$ which is precisely the definition of Sj{\protect\"o}gren and Kollmann.}, and calculated the r.m.s. displacement in a different way to take into account our peculiar way of averaging \refpar[eq:calcavg]. We also extend this discussion, in the rest of this section, to the case of finite systems.

\subsection{The ballistic regime (regime I)}
\label{sec:regimeIdiscussion}

 Since the small time behavior of each mode in \refpar[eq:solcorrel] is $(k_B T / N m)t^2$, and there are $N$ equivalent contributions to the sum we get
\beq
\langle \Delta x^2 (t)\rangle  \stackrel{t \to 0}{\sim}  { k_B T\over  m}t^2.
\label{eq:solcorreltpscourt}
\eeq
This result is independent of $N$, and thus valid in the thermodynamic limit too. Because of the inertial term in the Langevin equation \eqref{eq:chain}, at very small time each particle behaves independently from the others and undergoes ballistic flight at her thermal velocity $\sqrt{k_B T/m}$. 

In order to discuss the duration $\tau_{\rm ball}$ of this regime, let us assume a finite, but large (in a sense to be precised later) particle number $N$. The time evolution of $\langle \Delta x^2 (t)\rangle$ is determined by the mode $q = 0$ and a summation on all modes $q \neq 0$. All modes in the summation \eqref{eq:solcorrel}, hence the sum itself, behaves as $t^2$ on a timescale such that
\beq
t \leq \tau_{\rm ball} \equiv \hbox{min}\left(\frac{2}{\gamma}, \frac{1}{\sqrt{\omega_\pi^2 - \gamma^2/4}},\frac{\gamma}{\omega_\pi^2}\right).
\label{eq:tundeux}
\eeq

For weakly interacting or equivalently low density systems $\tau_{\rm ball} \approx 2/\gamma$, which was already heuristically derived in \eqref{t1weak}, and observed in Fig.~\ref{tlin}--a). For strongly interacting or equivalently high density systems, we get $\tau_{\rm ball} \approx 1/\omega_\pi \propto \sqrt{m/K} \propto \sqrt{m \kappa_T/\rho} $,  in perfect agreement with our observations Fig.~\ref{tlin}--b) and the heuristic derivations \eqref{eq:diffeffect} and \eqref{t1strong}.

\subsection{The collective regime (regime III)}
\label{sec:regimeIIIdiscussion}

This collective regime (regime III of Fig.~\ref{regimes}, see section~\ref{sec:regimeIIIdata}) is a property of finite systems only. This asymptotic behavior may be easily deduced from the sum \refpar[eq:solcorrel], which is dominated by the contribution of the mode $q = 0$ that scales as $[2 k_B T /(N m \gamma)] \times t$. This corresponds to the free diffusion of a particle of mass $N m$ in a thermal bath at temperature $T$. At very large time, the particles  are completely correlated and behaves as a single particle of effective mass the sum of all masses. The same result has been obtained in \cite{vanBeijeren83}. 

It is not difficult to estimate the time $\tau_{\rm coll}$. It is the time at which the contribution of the mode $q = 0$ dominates the sum of the contributions of all $N - 1$ other modes. Let us use the simplifying Debye approximation $\omega_q^2 = (K/m)q^2$. We thus get
\beq
\frac{2 k_B T}{N m \gamma}\tau_{\rm coll} \sim \sum\limits_{q \neq 0}\frac{2 k_B T}{N m \omega_q^2} \sim \frac{2 N k_B T}{4 \pi^2 K}\sum\limits_{i = 1}^{(N-1)/2}\frac{1}{i^2} \qquad \Longrightarrow \qquad \tau_{\rm coll} \sim \frac{N^2 m \gamma}{12 K},
\label{eq:tpslong}
\eeq
where we have used the fact that the sum is the generalized harmonic number $H_{(N-1)/2,2}$ which is equal to $\pi^2/6$ up to corrections of order $1/N$. The fact that this time scales as $N^2$ explains why this long time regime is seldom seen in simulations (to our knowledge, the only exception is \cite{Nelissen07}) or experiments. This is clearly illustrated by our Fig.~\ref{variances}--b), where we show that increasing $N$ shifts the long time regime toward longer times. This expression of $\tau_{\rm coll}$ is equal to our previous heuristic estimate \refpar[t2form]. It means that the reasoning at the basis of the derivation of eqn.~\eqref{eq:tpslong} includes in the right way the physical origin of the long time collective behavior of the finite chain.

As was already quoted in section~\ref{sec:regimeIIIdata}, at very small damping constant it is possible to observe at large time an evolution $\langle \Delta x^2 (t)\rangle = H_N t^2$. This is possible if the modes $q \neq 0$ are saturated while the mode $q = 0$ still evolves as $[ k_B T/(Nm)]t^2$. This requires $t < 1/\gamma$ (otherwise the mode $q = 0$ scales as $t$), and that the contribution of the $q\neq 0$ modes to the sum in \refpar[eq:solcorrel] be less than that of the mode $q = 0$.  Roughly speaking, we get
\beq
\left.\left\langle \Delta x^2 \left(t = \frac{1}{\gamma}\right)\right\rangle\right\vert_{q = 0} \sim \frac{k_B T}{N m}\left(\frac{1}{\gamma}\right)^2 > \sum\limits_{q \neq 0}\frac{2 k_B T}{N m \omega_q^2} \sim \frac{N m k_B T }{12 K} \quad\Longrightarrow\quad \gamma^2 < \frac{12 K}{N^2 m},
\label{eq:tpslongbis}
\eeq
where we have used the Debye approximation to estimate the contribution of the $q\neq 0$ modes. This means that this regime is to be observed at small damping $\gamma$ and small particles number $N$, which is precisely the case in Fig.~\ref{variances}--a) and Fig.~\ref{variances}--b). In this case, the estimate~\eqref{eq:tpslong} should be replaced by
\beq
\frac{k_B T}{N m} \tau_{\rm coll}^2 \sim \sum\limits_{q \neq 0}\frac{2 k_B T}{N m \omega_q^2} \sim \frac{2 N k_B T}{4 \pi^2 K}\sum\limits_{i = 1}^{(N-1)/2}\frac{1}{i^2} \qquad \Longrightarrow \qquad \tau_{\rm coll} \sim N\sqrt{\frac{ m }{6 K}}.
\label{eq:tpslongbis}
\eeq
This expression of $\tau_{\rm coll}$ is equal to our previous heuristic estimate \refpar[t2form2], allowing us to interpret the $t^2$ scaling at long time, in systems of few particles with small damping, as a collective behavior linked to the translationally invariant mode. This regime takes place when the greatest saturation time of the modes $q \neq 0$ is smaller than the time above which the mode $q = 0$ evolves as $t$ rather than $t^2$. The behavior $\langle \Delta x^2 (t)\rangle = H_N t^2$ may thus be observable when
\beq
\frac{1}{\omega_{2\pi/N}} < \frac{1}{\gamma} \qquad \Longrightarrow \qquad \frac{N}{2\pi}\sqrt{\frac{\gamma^2 m \kappa_T}{\rho}} < 1.
\label{eq:criterium}
\eeq
The relevance of this estimate is proved by Fig.\ref{tlin}--c) and Fig.\ref{tlin}--d). It also show that the time $\tau_{\rm lin}$ introduced in section~\ref{sec:discussion} is equal to $1/\gamma$.

\begin{figure}[htb]
\centering
\includegraphics[width=7cm]{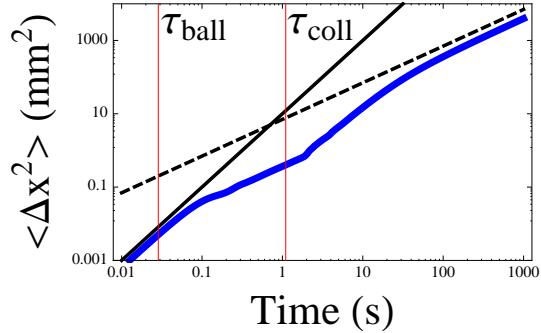}
\caption{\label{fig:fig5amath} Plot of $\langle\Delta x(t)^2 >$ (in mm$^2$) according to the time (in s.), in logarithmic scale, calculated from the analytical solution~\eqref{eq:solcorrel} for a density  $\rho = 533$~m$^{-1}$, $N = 32$, $T=10^{12}$~K, $\Gamma \approx 6.8$ and $\gamma = 0.1$s$^{-1}$. The solid line is of slope 2, the dashed line of slope 1. This plot is to be compared to the relevant plot in Fig.~\ref{variances}--a). We have indicated the times $\tau_{\rm ball}$ and $\tau_{\rm coll}$ respectively given by \eqref{eq:tundeux} and \eqref{eq:tpslongbis}. We see that both equation provides lower bounds for $\tau_{\rm coll}$.}
\end{figure}

\subsection{The correlated regime (regime II)}
\label{sec:regimeIIdiscussion}

Let us now discuss the intermediate regime~\footnote{In simulations of overdamped Langevin equation, for which the ballistic regime cannot be seen,  this intermediate regime is the first to be observed \cite{Herrera07,Herrera08,Henseler08,Centres10}.} which, between the individual ballistic regime (regime I) and the collective behavior (regime III), exhibit the correlated behavior of the particles.  While at asymptotically large time, all modes with finite (non zero) wave numbers have reached a constant value, in the intermediate regime, the physical behavior at a given time $t$ of the chain results from a subtle balance between the modes that are already saturated and those that still evolve. To simplify somewhat the discussion, we will considere the limit $\gamma \ll 2 \omega_{2\pi/N}$ when all modes are oscillating (underdamped) and the limit $\gamma \gg 2 \omega_\pi$, when all modes are overdamped.

\begin{figure}[htb]
\centering
\includegraphics[width=7cm]{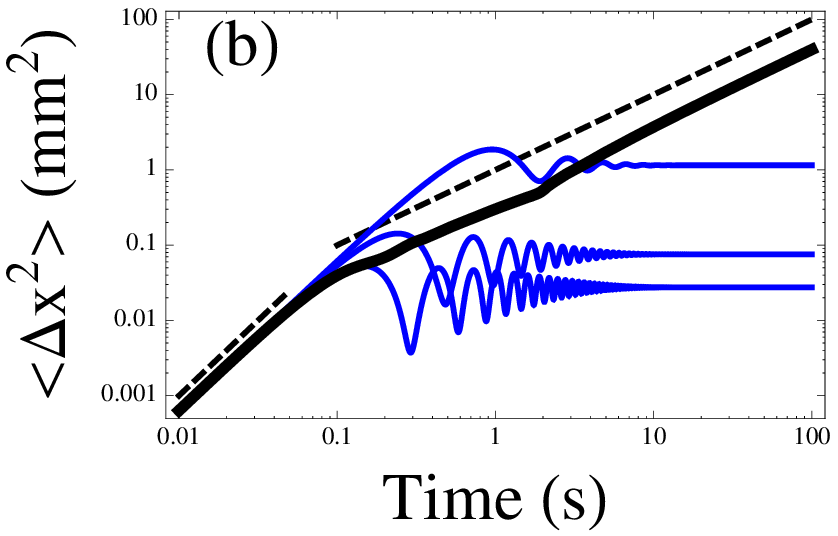}
\includegraphics[width=7cm]{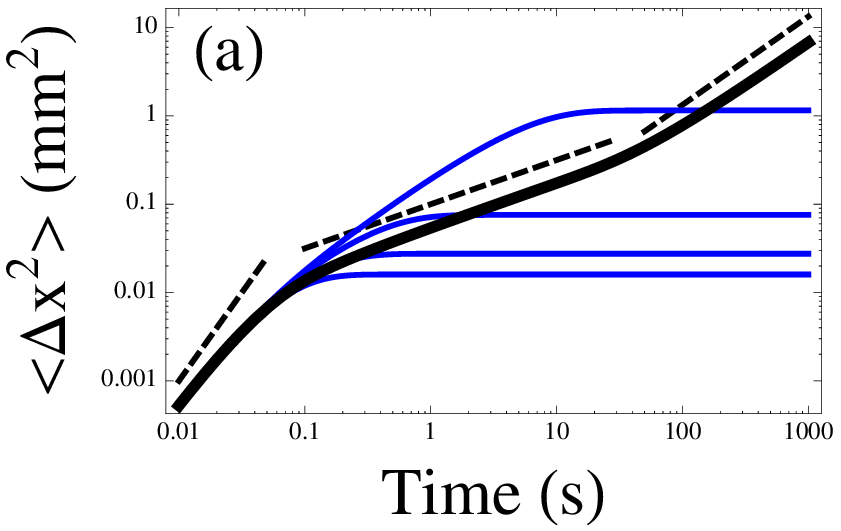}
\caption{\label{fig:modes} (Color online) Thick solid line~: plot of $\langle\Delta x(t)^2 >$ (in mm$^2$) according to the time (in s.), in logarithmic scale, for a density  $\rho = 533$~m$^{-1}$, $N = 32$, $T=10^{12}$~K, $\Gamma \approx 6.8$, calculated from \eqref{eq:solcorrel} for the relevant damping constant. (a) Underdamped regime, $\gamma = 1$~s$^{-1}$. The blue (gray) curves are, from top to bottom, the modes $q = \pi/16, q = \pi/4, q = 7\pi/16$. The dashed lines are, from left to right, of slopes 2 and 1. (b) Overdamped regime, $\gamma = 60$~s$^{-1}$. The blue (gray) curves are, from top to bottom, the modes $q = \pi/16, q = \pi/4, q = 7\pi/16, q = 5\pi/8$. The dashed lines are, from left to right, of slopes 2, $1/2$ and 1.}
\end{figure}

Let us first assume a very low damping. The modes oscillates until they reach a stationary value. The time evolution of $\langle \Delta x^2 (t)\rangle$ is due to the progressive disappearance of the contributions of the first oscillation of those modes $q \neq 0$ that have reached their maximum value. In Fig.~\ref{fig:modes}--a) this is graphically illustrated  with several underdamped modes ($\gamma/2 = 0.05$~s$^{-1} \ll \omega_\pi = 25$~s$^{-1}$), together with the complete sum~\eqref{eq:solcorrel}. As a first approximation,
\beq
\frac{1}{N^2}\left\langle \Delta X^2(q,t) \right\rangle \sim {2 k_B T \over N m  \omega_q^2}\left[1 - e^{-\gamma t/2} \cos\omega(q) t\right] \sim  { k_B T \over N m}t^2.
\label{eq:appmodelowgamma}
\eeq
At a given time $t$, the sum is dominated by the contributions of the modes that have not reached their first maximum, that is those modes such that $t < 1/\omega_q$. Let $n(t)$ be the number of such modes. In the Debye approximation $\omega_q = q\sqrt{K/m}$, the maximum wave number of those modes is $1/t\sqrt{K/m}$ so that an estimate of $n(t)$ is given by $n(t) \sim 2(N/2\pi)(1/t\sqrt{K/m})$ (the factor 2 takes into accounts the modes $\pm \vert q\vert$). The variance may thus be estimated as
\beq
\langle \Delta x^2(t) \rangle \sim { k_B T \over N m}t^2 \times n(t) \sim { k_B T \over \pi \sqrt{ m K}}t.
\label{eq:appvariancelowgamma}
\eeq
This is a normal diffusion with a diffusivity $k_B T /( 2\pi \sqrt{ m K})$ that depends on the stiffness of the interaction $K$, showing that it is a collective effect. This is in agreement with our observations, see Fig.~\ref{prefactors}--c). 

In the opposite limit of a very strong damping, all modes evolve monotonously toward their saturation value. As a first approximation,
\beq
\frac{1}{N^2}\left\langle \Delta X^2(q,t) \right\rangle \sim {2 k_B T \over N m  \omega_q^2}\left\{1 + \frac{\omega_-(q)[1 - \omega_+(q) t]}{\omega_+(q) - \omega_-(q)} \right\} \sim  { 2 k_B T \over N m \gamma}t,
\label{eq:appmodehighgamma}
\eeq
where we have used $\omega_+(q) - \omega_-(q) \approx \gamma$, $\omega_+(q) \approx \omega_q^2/\gamma$ and $\omega_-(q) \approx -\gamma$. A mode is saturated at a time $t > \gamma/\omega_q^2$. At a given time $t$, in the Debye approximation, the modes that increase with time are such that $q < \sqrt{m \gamma/(K t)}$. These reasoning is graphically illustrated in Fig.~\ref{fig:modes}--a) where we show several overdamped modes ($\gamma/2 = 30$~s$^{-1} > \omega_\pi = 25$~s$^{-1}$), together with the complete sum~\eqref{eq:solcorrel}. Their number is thus $n(t) \sim 2(N/2\pi)\sqrt{m \gamma/(K t)}$. The contributions of all such modes thus gives
\beq
\langle \Delta x^2(t) \rangle \sim { 2 k_B T \over N m \gamma}t \times n(t) \sim { 2 k_B T \over N m \gamma}t \times \frac{2 N}{2\pi}\sqrt{\frac{m \gamma}{K t}} = \frac{2 k_B T }{\pi \sqrt{m K \gamma}} t^{1/2} =  \frac{2 k_B T }{\pi} \sqrt{\frac{\kappa_T}{m \rho \gamma}} t^{1/2},
\label{eq:appvariancehighgamma}
\eeq
where we have introduced the isothermal compressibility $\kappa_T$ in the last expression to ease the comparison with the exact expression of $F_S$~\eqref{eq:mobform}. Taking into account the crudeness of our approximations, this estimate is extremely satisfactory because we recover the SFD behavior $\langle \Delta x^2(t) \rangle \propto t^{1/2}$, with a prefactor that is almost the exact one. 

In the general case, the modes with wave number $q$ such that $\gamma < 2\omega_q$ contribute to a $t$ scaling of the \msd, whereas the modes such that $\gamma > 2\omega_q$ contribute to the SFD behavior, that is a $t^{1/2}$ scaling of the \msd The typical time $\tau_{\rm sub}$ at which the \emph{subdiffusive} SFD behavior takes place is thus the inverse of this cut-off frequency, $\tau_{\rm sub} = 2/\gamma$. At a given particle number $N$, the minimum nonzero frequency is $\omega_{2\pi/N}$. If the damping constant is so small that all modes are underdamped ($\gamma < 2\omega_{2\pi/N}$) , the \msd scales as $t$ in the collective regime II. Increasing the damping, at fixed $N$, amounts to increase the number of overdamped modes, and favors the subdiffusive $t^{1/2}$ scaling for the \msd This is exemplified by Fig.~\ref{variances}--a). Increasing the particle number $N$, at fixed $\gamma$, amounts to decrease the frequency $\omega_{2\pi/N}$, hence to increase the number of overdamped modes, and favors the subdiffusive $t^{1/2}$ scaling for the \msd This is exemplified by Fig.~\ref{variances}--b).

The result \eqref{eq:appvariancehighgamma} is only approximate, so that the numerical prefactor cannot be trusted, but it explains under which conditions the SFD regime may be seen in finite systems with periodic boundary conditions [we remind the reader that this latter is the key assumption leading to \eqref{eq:mode}]. In appendix~\ref{sec:asympt}, we prove that the expression \eqref{eq:mobform} for the mobility of long ranged interacting systems is valid in the underdamped case $\gamma < 2\omega_\pi$ too, in the thermodynamic limit. 

\section{Conclusion}
\label{sec:conclusion}

In this paper, we study the SFD of a chain of particles with long ranged interactions, without the simplifying assumption of overdamped dynamics. We have focused our discussion on finite size effects and the influence of low damping. We use numerical simulations of the Langevin equation with the Gillespie algorithm \cite{Gillespie96AJP,Gillespie96PRE} and modelize the system as a chain of linear springs (spring constant $K$) and point masses ($m$) in a thermal bath at temperature $T$.

In our simulations data, we have identified several regimes for the time evolution of the mean square displacement (\msd ) $\langle \Delta x(t)^2\rangle$. At small times ($0 \leq t \leq \tau_{\rm ball}$), it evolves as $\langle \Delta x(t)^2\rangle = (k_B T/m) t^2$. This is a ballistic flight that traces back to the inertial effects, and is observed whatever the damping $\gamma$ or the particle number $N$. We recover this behavior in the thermodynamic limit ($N \to \infty$ at finite density) from our model. The prefactor of the $t^2$ scaling measured in our simulations is in excellent agreement with the theory.

For finite systems with periodic boundary conditions, an intermediate regime ($ \tau_{\rm ball} \leq t \leq \tau_{\rm coll}$) takes place. Depending on the respective values of the damping constant and the number of particles, we may observe either a diffusive behavior, a SFD behavior or successively both. We provide a physical explanation of those observations when we express the motion of the chain in terms of normal modes of oscillations. The \msd $\langle \Delta x(t)^2\rangle$ of the chain results from the superposition of all those modes. The mode associated to the null wave number is always overdamped, and similar to the motion of a free particle in a thermal bath, since no restoring force is exerted on it. The modes of finite (non zero) wave numbers have the same dynamics as an oscillator in an harmonic well. At long time, the \msd of all modes with non zero wave numbers saturate toward a constant value. The overdamped modes, which do not oscillate until they saturate, contribute to the SFD scaling $\langle \Delta x(t)^2\rangle \propto t^{1/2}$. The underdamped modes oscillate before their saturation, and contribute to a linear scaling $\langle \Delta x(t)^2\rangle \propto t$. 

In the thermodynamic limit, for systems of infinite number of particles, we exhibit analytically the  SFD behavior $\langle \Delta x(t)^2\rangle = F_S t^{1/2}$ at asymptotically large time. We recover the mobility $F_S$ that was previously calculated for long ranged interactions and overdamped dynamics \cite{Kollmann03,Sjogren07}, thus extending the previous calculation to systems with arbitrary damping. 

At asymptotically long time ($ t \gg \tau_{\rm coll}$), for a finite number of particles with periodic boundary conditions, the system behaves as an effective particle of mass $N \times m$. The physical origin of this behavior is the motion of the collective mode of null wave number, which is linked to the translational invariance of the system. For $t \geq \tau_{\rm lin} \geq \tau_{\rm coll}$, the system undergoes a linear diffusion with a diffusion coefficient $D_N = 2 k_B T/(N m \gamma)$. We show that this regime takes place whatever the value of $\gamma$. The duration of our simulations allows us to see this regime, and the measured diffusivity is in excellent agreement with its predicted value. We provide estimates of the time $\tau_{\rm coll} \sim N^2 m \gamma/K$ at large damping and $\tau_{\rm lin}\sim 1/\gamma$ at small damping. Those estimates are in good agreement with our simulations data. At small particle number and small damping, a new regime takes place at times $\tau_{\rm lin} \geq t \geq \tau_{\rm coll}$. It corresponds to the ballistic flight of the effective particle of mass $N \times m$, with $\langle  \Delta x(t)^2\rangle = (k_B T/N m)t^2$. In this case, the time $\tau_{\rm coll} \sim N\sqrt{m/K}$.

\acknowledgements

We thank J. Mokhtar  and F. van Wijland for helpful discussions.

\appendix
\section{Averaging}
\label{sec:average}

In this appendix, we calculate the averages used in our analysis of the simulations data. As explained in section~\ref{sec:algorithm} [see eqn.~\refpar[eq:calcavg]], we perform a double averaging. The first averaging is  ensemble averaging, done on the statistical distribution of the random force $\mu(t)$, and will be noted in this appendix $\langle \cdot\rangle_e$ for the sake of clarity. This averaging is involved in eqns.~\refpar[eq:wgn1], \refpar[eq:wgn2] and \refpar[eq:wgn4nodim]. The second averaging is performed on the statistical distributions of $X^0$ and $\dot X^0$ and will be denoted $\langle \cdot\rangle_0$. It is obvious that those two averaging operations commute, and that $\langle X_\mu (t)\rangle_0 = X_\mu (t)$ and $\langle X_d (t)\rangle_e = X_d (t)$. In \refpar[eq:defrmsFourier], the ensemble averaging is made on the random force $\mu(t)$ so that
\beq
\langle \Delta X^2(q,t) \rangle_e \equiv \left\langle \bigl[X(q,t) - \langle X(q,t)\rangle_e\bigr]\bigl[X(-q,t) - \langle X(-q,t)\rangle_e\bigr]\right\rangle_e,
\label{eq:defrmsFouriermod}
\eeq

\subsection{The mode $q = 0$}
\label{sec:mode0}

For the translational invariant mode $q = 0$, all trajectories beginning at a given time $t_0$ are equivalent so that all initial positions $X^0$ are equivalent. It is easy to check that $\langle\langle X(t) - X^0 \rangle_e\rangle_0 = \langle X_d(t) \rangle_0 + \langle X_\mu(t) \rangle_e = 0$. For the same reason, the double average $\langle\langle X_d(t)  X_\mu(t) \rangle_e\rangle_0 = 0 $ because this term is linear in $\dot X^0$ and in $\mu(0,t)$. The variance is thus
\beq
\langle\langle \Delta X^2 \rangle_e\rangle_0 = \left\langle\left\langle \left[X(t) - X^0\right]^2 \right\rangle_e\right\rangle_0 = \left\langle \left[X_d(t) - X^0\right]^2 \right\rangle_0 + \left\langle X_\mu(t)^2 \right\rangle_e
\label{eq:defvariance}
\eeq

The variance for the deterministic part of the displacement is
\beq
\langle\langle \Delta X_d^2(q,t) \rangle_e\rangle_0 =  \frac{\left\langle\vert\dot X^0\vert^2\right\rangle_0}{ \gamma^2}\left[1 - e^{-\gamma t}\right]^2 = \frac{N k_B T}{m \gamma^2}\left[1 - e^{-\gamma t}\right]^2,
\label{eq:modezerodeterm}
\eeq
where the last expression is provided by the equipartition theorem for the potential energy.

We then have to calculate the variance for the fluctuating part, $\left\langle X_\mu(t)^2 \right\rangle_e$. We begin by calculating the following time derivative,
\beqa
\frac{\dd}{\dd t} \langle \Delta X_\mu^2 \rangle_e &=& \frac{\dd}{\dd t} \langle X_\mu(t)X_\mu(t) \rangle_e \\ 
\label{eq:deriv}
&=& \langle X_\mu(t)\dot X_\mu(t) \rangle_e + \langle \dot X_\mu(t)X_\mu(t) \rangle_e = 2{\rm Re}\langle X_\mu(t) \dot X_\mu(t) \rangle_e
\nonumber
\eeqa
From \refpar[eq:modezerosolu], we get
\beq
\langle X_\mu(t) \dot X_\mu(t) \rangle_e = \frac{1}{m^2}\int\limits_0^t \dd t' \int\limits_0^{t'} \dd t'' e^{-\gamma(t' - t'')}\int\limits_0^t \dd t''' e^{-\gamma(t - t''')} \langle\mu(q = 0,t'')\mu(q = 0,t''')\rangle_e.
\label{eq:calcint}
\eeq
The correlation for the random noise is $\langle \mu(l,t)\mu(l',t')\rangle_e = g \delta_{l l'}\delta(t - t')$, with $g = 2 m k_B T$. Besides, $\mu(q = 0,t) = \sum_{l = 1}^N \mu(l,t)$, hence
\beq
 \langle\mu(q = 0,t)\mu(q = 0,t')\rangle_e = \sum\limits_{l = 1}^N\sum\limits_{l' = 1}^N \langle\mu(l,t)\mu(l',t')\rangle_e = 2 N m k_B T \delta(t - t').
\label{eq:correlnoisefourier}
\eeq
Performing the integrations in \refpar[eq:calcint], we get
\beq
\frac{\dd}{\dd t} \langle \Delta X_\mu^2 \rangle_e = 2\frac{N k_B T}{m \gamma}\left(1 + e^{-2\gamma t} - 2 e^{-\gamma t}\right),
\label{eq:derivsolu}
\eeq
and a last integration gives the final result
\beq
\langle \Delta X_\mu^2 \rangle_e = 2\frac{N k_B T}{m \gamma}\left(t  + \frac{1 -  e^{-2\gamma t}}{2\gamma} -2 \frac{1 -  e^{-\gamma t}}{\gamma}\right).
\label{eq:randomfree}
\eeq
Injecting this result and \refpar[eq:modezerodeterm] in \refpar[eq:defvariance] gives the final result \refpar[eq:doubleavgzero] as stated in the text.

\subsection{The modes $q \neq 0$}
\label{sec:modenon0}

Physically, the dynamics of a mode $X(q,t)$ with $q \neq 0$ is identical to the motion of a damped harmonic oscillator \refpar[eq:mode]. In this case, the initial values $X(q,0)$ correspond to an initial position in a potential well and are thus not equivalent. On the other hand, the stationary state is quickly reached and the statistical distribution on $X(q,0)$ is readily described taking different trajectories. In the data analysis, a trajectory  is given by looking at values $X(q,t + t_0) - X(q,t_0)$, and varying the time $t_0$ amount to varying the initial value $X(q,0)$. The expression \refpar[eq:defrmsFourier] of the variance is thus replaced by
\beqa
\langle\langle \Delta X^2(q,t) \rangle_e\rangle_0 &\equiv& \left\langle \left\langle \bigl[X(q,t) - X(q,0) -\langle\langle X(q,t)-X(q,0)\rangle_e\rangle_0\bigr]\right.\right. \times \nonumber\\ & \times & \left.\left.\bigl[X(-q,t) - X(-q,0)- \langle\langle X(-q,t) - X(q,0)\rangle_e\rangle_0\bigr]\right\rangle_e\right\rangle_0,
\label{eq:defrmsFourierbis}
\eeqa
As shown by \refpar[eq:solcorrel], $X(q,t) = X_d(q,t) + X_\mu(q,t)$ where $X_d(q,t)$ is the deterministic part, linear in $X(q,0)$ and $\dot X(q,0)$, and $X_\mu(q,t)$ the random part. It is not necessary to give explicitly the random part, all we need to know is that it is linear in the random force $\mu(q,t)$. Since $\langle X(q,0) \rangle_0 = 0$, $\langle \dot X(q,0) \rangle_0 = 0$ and $\langle X_\mu(q,t) \rangle_e = 0$, we have $\langle\langle X(q,t) - X( q,0) \rangle_e\rangle_0 = 0$. The cumbersome expression \refpar[eq:defrmsFourierbis] may thus be simplified to give
\beqa
\langle\langle \Delta X^2(q,t) \rangle_e\rangle_0 & = &  \left\langle \left\langle \bigl[X(q,t) - X(q,0)\bigr]\bigl[X(-q,t) - X(-q,0)\bigr]\right\rangle_e\right\rangle_0 \nonumber \\
& = & \left\langle \left\langle X(q,t) X(-q,t) \right\rangle_e\right\rangle_0 + \left\langle \left\langle  X(q,0) X(-q,0)\right\rangle_e\right\rangle_0 -  \\ & \quad & \qquad \qquad \qquad - 2\hbox{Re}\left[\left\langle \left\langle X(q,t) X(-q,0)\right\rangle_e\right\rangle_0\right] \nonumber
 \label{eq:defrmsFourierbissimp} 
\eeqa
This may be simplified further, because the stationarity implies $\left\langle \left\langle X(q,t) X(-q,t) \right\rangle_e\right\rangle_0  =  \left\langle \left\langle  X(q,0) X(-q,0)\right\rangle_e\right\rangle_0 = N k_B T/K$. Moreover, since $X(-q,0)$ do not depends on $\mu$ we have $\left\langle \left\langle X(q,t) X(-q,0)\right\rangle_e\right\rangle_0 =  \left\langle X_d(q,t) X(-q,0)\right\rangle_0$. The averaging is thus easily performed to give \refpar[eq:flucqnonnul].

\section{The chain of springs in the thermodynamic limit}
\label{sec:asympt}

In the thermodynamic limit $N \to \infty$, the discrete sum in \refpar[eq:solcorrel] may be replaced by an integral (taking advantage of the fact that the expressions for $q \neq 0$ are valid in the limit $q \to 0$),
\beq
\langle \Delta x^2 (t)\rangle = {2 k_B T\over m} {1 \over \pi}\int\limits_0^\pi \dd q {1 \over \omega_q^2}\left[1 + {\omega_-(q)e^{\omega_+(q)t} \over \omega_+(q) - \omega_-(q)} - {\omega_+(q)e^{\omega_-(q)t} \over \omega_+(q) - \omega_-(q)}\right],
\label{eq:solcorrelinteg}
\eeq
where we used the rule $(1/N)\sum_q \longrightarrow (1/2\pi)\int_{-\pi}^\pi$ and the invariance $q \to -q$. 

In the limit $\gamma \gg 2\omega_\pi$, this integral may be expressed in closed form \cite{Sjogren07}. This is not possible in the general case, but we may obtain its asymptotic behavior at large time using Laplace method \cite{Nayfeh93}. To this end, we express the time derivative
\beq
{\partial \langle \Delta x^2 (t)\rangle \over \partial t} = {2 k_B T\over m} {1 \over \pi}\int\limits_0^\pi \dd q {e^{\omega_+(q)t} - e^{\omega_-(q)t} \over \omega_+(q) - \omega_-(q)}.
\label{eq:derivinteg}
\eeq
For underdamped modes, $\omega_\pm = -\gamma/2 \pm i \sqrt{\omega_q^2 - \gamma^2/4}$, the large time behavior is dominated by $\exp[-\gamma t /2]$. Since $\lim\limits_{q \to 0}\omega_q = 0$, the modes with small wave numbers are always overdamped. The asymptotic behavior of the integral is dominated by the neighborhood of the maximum of $\omega_+(q) $ at $q = 0$, with $\omega_+(0) = 0$, $\omega_+'(0) = 0$ and $\omega_+''(0) = -2K/(m \gamma)$. The leading term in \refpar[eq:derivinteg] is thus \cite{Nayfeh93}
\beq
{\partial \langle \Delta x^2 (t)\rangle \over \partial t} \stackrel{t \to \infty}{\sim} {2 k_B T\over \pi m} {1\over 2}\left[{2 \over - t \omega_+''(0)}\right]^{1/2}{1\over \gamma}e^{\omega_+(0)t}\Gamma\left({1\over 2}\right) = {2} \left[{(k_B T)^2 \over \pi m \gamma K}\right]^{1/2}{1\over 2 t^{1/2}}.
\label{eq:devSFDderiv}
\eeq
The leading behavior of the variance, at large time, is thus
\beq
 \langle \Delta x^2 (t)\rangle  \stackrel{t \to \infty}{\sim}   2\left({k_B T D_0 \kappa_T \over \pi \rho}\right)^{1/2} t^{1/2}.
\label{eq:devSFDderiv}
\eeq
To get the last expression, we have used the relation $m \gamma D_0 = k_B T$ and introduced the isothermal compressibility of the chain, which is $\kappa_T = \rho/K$ \cite{Brillouin53}.

We recover the behavior for single file diffusion of particles interacting with a long-ranged potential, as was shown by Kollmann \cite{Kollmann03} (see \cite{Coste10} for a rewriting of Kollmann's result in terms of the isothermal compressibility) and Sj\"ogren \cite{Sjogren07}. Those previous calculations were done under the simplifying assumption of overdamped dynamics ($\gamma \gg 2\omega_\pi$). We nevertheless recover the same result because the asymptotics of integral \refpar[eq:derivinteg] is dominated by the modes of small wavenumbers $q \ll 1$. Whatever the finite value of the damping constant $\gamma$, they are always overdamped because $\omega_{q \to 0} = 0$.
\newpage

\bibliography{biblio}

\end{document}